%% file: main.tex
\documentclass[sn-mathphys,Numbered]{sn-jnl}

\usepackage{graphicx}%
\usepackage{multirow}%
\usepackage{amsmath,amssymb,amsfonts}%
\usepackage{amsthm}%
\usepackage{mathrsfs}%
\usepackage[title]{appendix}%
\usepackage[dvipsnames]{xcolor}
\usepackage{textcomp}%
\usepackage{manyfoot}%
\usepackage{booktabs}%
\usepackage{algorithm}%
\usepackage{algorithmicx}%
\usepackage{algpseudocode}%
\usepackage{listings}%
\usepackage{siunitx}%

\theoremstyle{thmstyleone}%

\theoremstyle{thmstyletwo}%

\theoremstyle{thmstylethree}%

\newcommand{\mb}[1]{\mathbf{#1}}

\begin{document}

\title[COLIS]{COLIS: an advanced light scattering apparatus for investigating the structure and dynamics of soft matter onboard the International Space Station}

\author[1]{\fnm{Alessandro} \sur{Martinelli}}\email{alessandro.martinelli@umontpellier.fr}
\author*[2]{\fnm{Stefano} \sur{Buzzaccaro}}\email{stefano.buzzaccaro@polimi.it}
\author[3]{\fnm{Quentin} \sur{Galand}}\email{Quentin.Galand@vub.be}
\author[1,4]{\fnm{Juliette} \sur{Behra}}\email{juliette.behra@instrumat.ch 
}
\author[5]{\fnm{Niel} \sur{Segers}}\email{niel.segers@redwirespace.eu}
\author[5]{\fnm{Erik} \sur{Leussink}}\email{erik.leussink@redwirespace.eu}
\author[5,6]{\fnm{Yadvender Singh}\sur{Dhillon}}\email{yadvendersingh.dhillon@hdr.mq.edu.au}
\author[3]{\fnm{Dominique} \sur{Maes}}\email{Dominique.Maes@vub.be}
\author[7]{\fnm{James} \sur{Lutsko}}\email{Jim.Lutsko@ulb.be}
\author[2]{\fnm{Roberto} \sur{Piazza}}\email{roberto.piazza@polimi.it}
\author*[1,8]{\fnm{Luca} \sur{Cipelletti}}\email{luca.cipelletti@umontpellier.fr}

\affil[1]{\orgdiv{L2C}, \orgname{Universit\'e Montpellier}, \orgaddress{\street{P. Bataillon}, \city{Montpellier}, \postcode{34095}, \country{France}}}
\affil[2]{\orgdiv{Department of Chemistry, Materials Science, and Chemical Engineering (CMIC)}, \orgname{Politecnico di Milano}, \orgaddress{\street{Piazza Leonardo da Vinci 32}, \city{Milano}, \postcode{20133}, \country{Italy}}}
\affil[3]{\orgdiv{Structural Biology Brussels}, \orgname{Vrije Universiteit Brussel}, \orgaddress{\street{Pleinlaan 2}, \city{Brussels}, \postcode{1050},  \country{Belgium}}}

\affil[4]{Present address: \orgname{Instrumat AG}, \orgaddress{\street{Chemin de la Rueyre 116/118}, \city{Renens}, \postcode{CH-1020}, \country{Switzerland}}}

\affil[5]{\orgdiv{Design \& Development}, \orgname{Redwire Space N. V.}, \orgaddress{\street{Hogenakkerhoekstraat 9}, \city{Kruibeke}, \postcode{9150}, \country{Belgium}}}

\affil[6]{Present address: \orgdiv{School of Engineering}, \orgname{Macquarie University}, \orgaddress{\street{Wallumattagal Campus, Macquarie Park}, \city{Sidney}, \postcode{NSW 2109}, \country{Australia}}}

\affil[7]{\orgdiv{Center for Nonlinear Phenomena and Complex Systems, CP231 and BLU-ULB Space Research Center}, \orgname{Universit\'{e} Libre de Bruxelles}, \orgaddress{\street{Boulevard du Triomphe}, \city{Brussels}, \postcode{1050},  \country{Belgium}}}

\affil[8]{\orgname{Institut Universitaire de France}, \orgaddress{\street{1, Rue Descartes}, \city{Paris}, \postcode{75231}, \country{France}}}

\abstract{Colloidal Solids (COLIS) is a state-of-the-art light scattering setup developed for experiments onboard the International Space Station (ISS). COLIS allows for probing the structure and dynamics of soft matter systems on a wide range of length scales, from a few nm %
to tens of microns, and on time scales from 100 ns to tens of hours. In addition to conventional static and dynamic light scattering, COLIS includes depolarized dynamic light scattering, a small-angle camera, photon correlation imaging, and optical manipulation of thermosensitive samples through an auxiliary near-infrared laser beam, thereby providing a unique platform for probing soft matter systems. We demonstrate COLIS through ground tests on standard Brownian suspensions, and on protein, colloidal glasses, and gel systems similar to those to be used in future ISS experiments.}

\keywords{light scattering, microgravity, International Space Station, colloids, proteins, gels, glasses}

\maketitle

Condensed soft matter systems comprise objects with sizes in the nanometer to several micrometers range suspended in a fluid, most often water. While the nature of the individual objects may vary, from solid colloidal particles to liquid droplets, surfactant molecules, polymers and proteins, and even cells and their constituents, soft systems share common features, such as the relevance of Brownian motion and the high susceptibility to external fields, even of modest strength, which originated the very term `soft matter'~\cite{brezin_demain_2009}. Soft systems are ubiquitous in everyday products and industry and are often investigated as model systems for their atomic or molecular condensed matter counterparts. Current challenges in soft matter include understanding and rationalizing out-of-equilibrium phenomena and addressing disorder and the existence of a hierarchy of length scales, due to the organization of the individual constituents in superstructures, e.g. in crystallization or gelation, or by self-assembly~\cite{barrat_soft_2024}.

While gravitational forces are rarely relevant at the single object scale, they do play a major role when, e.g., colloids or proteins assemble to form large structures, which may rapidly settle, or through which gravitational stress may be transmitted and accumulated over macroscopic distances. For some systems, the effect of gravity may be mitigated by matching the density of the background fluid to that of the suspended objects; however, buoyancy matching may be unfeasible without changing significantly the physicochemical properties of the system, or while concomitantly achieving near-refractive index matching, required for optical observations. Furthermore, shear forces due to convective currents are practically unavoidable in terrestrial gravity and can significantly impact growth processes. Soft systems are thus natural candidates for microgravity research~\cite{chaikin_grand_2021} and have been extensively studied in experiments in sounding rockets, or onboard the space shuttle and the International Space Station (ISS).  Crystallization experiments on macromolecules~\cite{Shuttle}, proteins~\cite{GARCIARUIZ2001149,zegers_counterdiffusion_2006,DM,patino-lopez_protein_2012}, drugs~\cite{reichert_pembrolizumab_2019}, and colloids~\cite{zhu1997crystallization, russel1997dendritic,cheng2001crystallization} have allowed for the characterization of growth rates and the obtention of crystals of larger size and better quality than on Earth. They have also unveiled unforeseen features, such as the dendritic growth of colloidal crystals~\cite{russel1997dendritic} and the emergence of crystalline order in dense colloidal suspensions that formed glasses in terrestrial gravity~\cite{zhu1997crystallization}. Colloidal gels have been investigated as prototypical network formers, unveiling slow restructuring~\cite{manley_timedependent_2005} and pointing to the role of gravity in limiting the growth of the fractal aggregates that ultimately form the gel~\cite{manley_limits_2004a}. In closely related fields, gravity has been shown to deeply affect the behaviour of nonequilibrium fluctuations in complex fluids~\cite{cerbino2015dynamic}, the stability of foams~\cite{galvani_hierarchical_2023}, and the behavior of granular matter~\cite{born_dense_2017,ozaki_granular_2023}, and synthetic and biological active matter~\cite{acres2021influence,volpe2022active}.

Optical methods are powerful and convenient tools to probe the structure and dynamics of soft matter. They are non-invasive and can be implemented in compact, lightweight apparatuses suitable for space flights, with designs based on low-magnification imaging and microscopy, see, e.g.,~\cite{wareing_flightscope_2024} and references therein, holography~\cite{kebbel_digital_1999}, shadowgraphy~\cite{braibanti_european_2019}, and light scattering~\cite{lant_physics_1997,born_soft_2021a} and differential dynamic microscopy~\cite{safari_differential_2017,mazzoni2013sodi}. Here, we describe COLIS, an advanced light scattering-based space instrument developed for the European Space Agency (ESA) Colloidal Solids project, aiming at investigating onboard the ISS the origin, formation and dynamics of colloidal and protein crystals, and colloidal glasses and gels. COLIS enables in-situ monitoring of the dynamics of physical processes during and after solidification, which is needed to assess the role played by gravity on the properties of growing structures, addressing long-standing questions such as the effect of gravity on protein nucleation~\cite{ProteinReview} and on anomalous, ultraslow and heterogeneous relaxations in gels and glasses~\cite{cipelletti2000universal,filiberti_multiscale_2019,philippe2018glass}. COLIS features an unprecedented combination of light scattering techniques that can be used simultaneously: conventional Dynamic Light Scattering (DLS)~\cite{berne76} at three angles, paired, at the same angles, by Photon Correlation Imaging (PCI)~\cite{duri2009resolving}, a method that blends scattering and imaging to probe heterogeneous dynamics; small-angle static and dynamic light scattering leveraging the multispeckle method~\cite{luca_cipelletti_ultralow-angle_1999} to address slow, non-ergodic dynamics, as in gels and glasses, and absorbance measurements. All scattering detectors are equipped with a polarizer, allowing for both conventional polarized light scattering and depolarized scattering, e.g. for probing rotational dynamics of optically anisotropic samples. Finally, COLIS affords advanced sample environment controls: fluid stirring, accurate temperature control including the possibility of performing fast $T$ jumps, and local heating of the scattering volume through a near-infrared auxiliary laser beam, for the optical manipulation of thermosensitive samples~\cite{Ruzzi_PhD}.  

\begin{figure}
\centering
\includegraphics[width=0.95\textwidth]{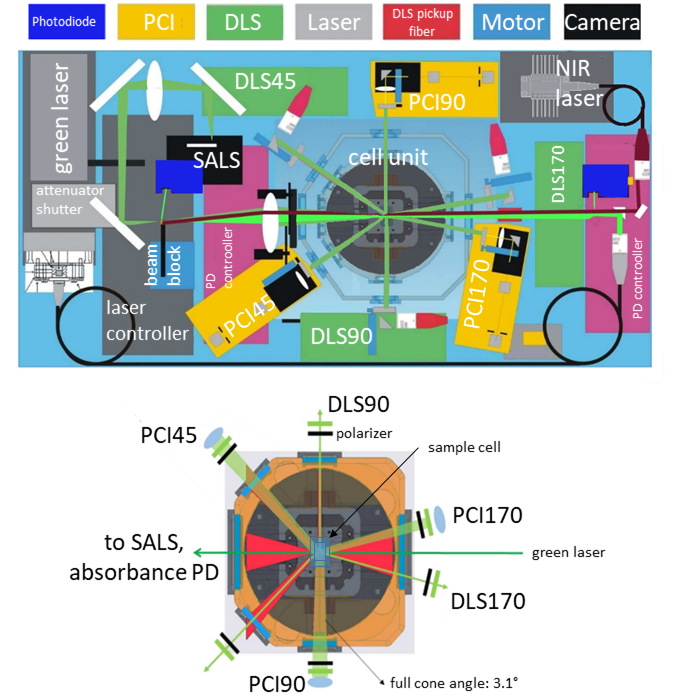}
\caption{Schematic view of the experimental unit and cell module of COLIS. Top panel: Experimental unit showing the optical diagnostic lines. Light green: optical path of the green laser ($\lambda_0 = 532$ nm); dark green:  paths of the scattered light collected by the optical diagnostic of the instrument; dark red: near infrared laser ($\lambda_0 = 980$ nm). The NIR and optical beams are colinear; here they have been slightly offset for clarity. Bottom panel; section of the cell module E, indicating the optical accesses for the various diagnostic.}
\label{fig:setup}
\end{figure}

 \section{The COLIS setup}\label{sec:COLIS}
COLIS is designed to be operated on board of the International Space Station (ISS), accommodated inside the Microgravity Science Glovebox (MSG) host facility. It comprises the following main diagnostics, see Fig.~\ref{fig:setup}: 
\begin{enumerate}
\item A power-adjustable (max 150 mW) \emph{optical laser} (wavelength 532 nm) equipped with a shutter. 
\item A \emph{near infrared (NIR) laser} (wavelength 980 nm) coaxial to the green laser, to dynamically heat water-based solutions. The laser intensity can be easily modulated, e.g. cyclically.
\item Three \emph{Dynamic Light Scattering} lines, at forward, 90° and backscattering angles, denoted by DLS45, DLS90 and DLS170 in the following. The scattered light is collected by collimated single mode fibers, all equipped with a rotating linear polarizer and connected to photon counters and hardware correlators.
\item Three \emph{Photon Correlation Imaging} lines (PCI45, PCI90, PCI170), at the same scattering angles as the DLS lines, and all equipped with rotating linear polarizers. For each PCI line, the scattered light is imaged on a temperature-stabilized CMOS camera for multispeckle analysis. 
\item A \emph{small angle light scattering (SALS)} setup, equipped with a beam block and a rotating linear polarizer. Scattered light at angles in the range 0.5° to 9° is recorded by a temperature-stabilized CMOS camera.

\end{enumerate}

COLIS adopts a modular design concept similar to the SODI experiment \cite{braibanti_european_2019}. It comprises three modules that will be separately uploaded in soft stowage bags and then integrated in the MSG by the ISS crew: 
\begin{enumerate}
\item \emph{Experiment Unit (EU)}, top of Fig.~\ref{fig:setup}. The EU contains all the optical sub-systems needed for the scattered light diagnostics (DLS, PCI, SALS). Its main function is the acquisition of the scientific data. 
\item \emph{Cell Module (CM)}, bottom of Fig.~\ref{fig:setup}. The CM contains the actual sample and can be inserted in a slot of the EU. The CM provides two levels of containment for the experiment liquid, the third one being that of the EU. The CM contains all the necessary hardware for the sample environment (stirring mechanism, Peltier elements, pressure compensator etc.). For each new experiment, a separate CM is uploaded and inserted into the EU by the ISS crew. 
\item \emph {Image Processing Unit (IPU)}. All the power and data lines of the MSG are connected to the IPU, which houses a powerful computer, that fulfils functions such as communicating with the outside world, controlling the experiment timeline and parameters, commanding the thermal control and sample stirring, pre-processing the PCI ans SALS images to reduce the amount of data to be downlinked to ground. These functionalities allow for quasi-autonomous operation for prolonged period of times, up to several days, and for uplinking new experimental protocols to fit future science needs. Two data disks are uploaded as part of the IPU. They are used to store the raw and pre-processed scientific data and can be exchanged on the ISS, if needed. They can be returned to ground, for further in-depth analysis of the raw data.
\end{enumerate}

\subsection{Overview of the Experimental Unit}\label{sec:overview_COLIS}

\subsubsection{Optical laser source}
The optical laser source is a continuous--wave diode--pumped laser operating at a fixed in-vacuo wavelength $\lambda_0 = $ 523 nm. The laser head is pigtailed and the the maximum beam power is $\ge 170~\si{mW}$ at the fiber exit and $\ge 150~\si{mW}$ on the sample. A  collimator  at the fiber exit generates a collimated beam with  a diameter of $1.5~\si{mm}  \pm 0.1~\si{mm}$, with a divergence angle of $0.013^{\circ}\pm 0.02^{\circ}$.  A laser shutter and attenuator allows for controlling the sample illumination. In case of shutter failure, an ISS crew member can manually set the shutter in the (permanently) open position.
For monitoring purpose, the laser light is sampled by a beam sampler  and detected by a low-noise photodiode.

\subsubsection{Near infrared laser source}
The NIR laser is a fiber-Bragg-grating and thermally-stabilized laser diode operating at $\lambda_0 = 975$ nm, with maximum power 500 mW.  The laser intensity is monitored by an internal photodiode. A collimator generates a collimated beam with a diameter of $2.10~\si{mm} \pm  0.11~\si{mm}$ with a divergence angle of $0.017^{\circ}$. The NIR beam is coaxial to the green laser beam, so as to  heat the scattering volume. After passing through the sample, the NIR beam is stopped by the SALS beam block. 

\subsubsection{Dynamic Light Scattering lines}
For DLS, pick-up fibers collect light scattered at three angles, whose values in air, with respect to the incident beam direction, are $45^{\circ}$, $90^{\circ}$, and $166.65^{\circ}$. The corresponding scattering angles $\theta$ in the sample depend on the solvent refractive index, due to refraction effects. They can be calculated using Snell's law, considering simply a solvent-air interface, since the cell glass windows are parallel slabs that do not change the propagation direction. For water-based samples, using $n_{H_2O} = 1.33$, one obtains $\theta = 32.12^{\circ}$, $90^{\circ}$, and $170^{\circ}$, respectively.

Each collecting fiber is equipped with its own polarizer (polarization extinction ratio $>$ 10000:1), to perform polarized and depolarized light scattering. The scattered light collected by the fibers is fed to three avalanche photodiodes, connected to hardware correlators, for real-time calculations of the  intensity correlation function for delay times from 12.5 ns to about one hour.

\subsubsection{Photon Correlation Imaging lines}
\label{sec:PCI_overview}
Each PCI line comprises optics that image the scattering volume, as seen from one of the three angles defined above for DLS, on a USB3.0 CMOS camera, see Fig.~\ref{fig:cameras} for typical PCI images. The PCI cameras have a resolution of 2048 $\times$ 1536 pixels and a pixel size of 3.45 $\mu m$ (sensor size: 7.1 mm $\times$ 5.3 mm). At 8-bit, the resolution used in COLIS, each camera has a maximum full-frame acquisition rate of 120 Hz. The frame rate may be reduced by decreasing the number of acquired pixel rows. For example, in order to image the full scattering volume the PCI90 camera is run at 2048 $\times$ 640 pixels, corresponding to a field of view of 5.1 mm $\times$ 1.6 mm. Under this conditions, the acquisition rate is up to 287 Hz. Each camera is cooled to room temperature, using a Peltier module. The imaging optics comprises an objective lens and a circular diaphragm placed in its focal plane. The purpose of the diaphragm is to control the size of the speckles formed on the CMOS detector (about 3 pixels), and to guarantee that the scattering angle and magnification of the light rays reaching the detector are uniform across the detector area (telecentric conditions). A magnification of 1.38 is chosen, resulting in a full-frame field of view of 5.1 mm $\times$ 3.8 mm. 

\subsubsection{Small angle light scattering line} 

The optical path of SALS may be divided into two parts: right after the cell module, a Fourier lens generates a far-field scattering pattern in its focal plane, where a beam block redirects the unscattered beam to a photodiode, for absorbance measurements. The beam block can be translated along two orthogonal directions in the focal plane, to optimize its positioning. The second part is composed of a relay lens that forms an image of the  plane of the beam block on the sensor of the SALS camera. The SALS optics is designed to collect light scattered at angles $\theta \le 9^\circ$. The speckle size is slightly angle—dependent, ranging from 2 to 5 pixels. The SALS line is equipped with a USB3.0 camera mounted on a Peltier cooler. A CMOS detector is used to avoid blooming, which is particularly important for small angle measurements, where the scattered intensity may vary over several orders of magnitude over the detector area.  The sensor has 2048 x 2048 pixels, with a pixel size of 5.5 $\si{\micro m}$. 
A polarizer (extinction rate $\ge$ 10000:1) is placed behind the beam block.  The polarizer can be rotated continuously in order to further control, for non—polarizing samples, the light intensity that reaches the camera, or to perform depolarized light scattering.

\subsubsection{Sources and image acquisition synchronization}

A TTL pulse generator with time resolution of 0.1 ms is used to control the timing of the image acquisition of the PCI and SALS cameras, the status of the NIR laser (ON/OFF) and the operation of the optical laser shutter. A trigger pattern usually lasting a few tens of seconds and comprising hundreds of image acquisitions is typically repeated cyclically, for an overall measurement duration up to tens of hours. The pulse generator guarantees that the camera images are taken sequentially, to avoid bus conflicts, and that the shutter is open while acquiring images. It is also used to illuminate cyclically the sample with the NIR beam, to create a periodic thermal perturbation.

\subsection{Cell module}

The cell module (CM) of COLIS is an exchangeable unit that contains the experiment fluid and all the hardware for thermal control and sample stirring, see Fig.~\ref{fig:setup}. The CM provides: 
\begin{itemize}
\item \emph{Optical access} for the optical and NIR laser beams and for all scattered light measurement lines. 
\item \emph{Control and monitoring} of the cell temperature.
\item \emph{Two levels of containment} of the experiment fluid, through the sample optical cell in the center of the CM and the six CM windows. 
\item \emph{Stirring} of the experiment fluid through a magnetic bar placed in the cell and actuated by a rotating magnet placed below the cell.
\end {itemize}

Two different different types of Cell Modules are designed: CM E and D.

\subsubsection{Cell module E}
The CM E is designed to ensure low thermal gradients and high stability of the experiment fluid, to allow for experiments lasting up to several days at a fixed temperature.  Its thermal inertia is still low enough to allow for imposing several kinds of thermal histories, such as temperature jumps (upward $T$ jump of 5°C with a rate up to 5°C/min), ramps, and periodic temperature oscillations (e.g. a nearly sinusoidal $T$ oscillation with period 60 sec and peak-to-peak amplitude of 2°C).
The sample temperature is monitored by 3 thermistors placed in thermal contact with the outside walls of the optical sample cell. The inner dimensions of the optical cell in the scattering plane (the plane of Fig.~\ref{fig:setup}) are: 5 mm along the direction of the optical axis, 9 mm in the direction orthogonal to the optical axis, to accomodate the DLS45 and TRC45 lines. The height of the cell is 7 mm.
The stirring system impose the rotation of a magnetic bar within the sample cell at up to 5000 rpm in both directions. It applies a torque sufficient to fully homogenize fluids with a viscosity of  maximum $100~\si{mPa s}$ (100 times the water viscosity) and a max yield stress of $5 ~\si{Pa}$. The 
stirrer design shall allow complete re-dispersion within 30 minutes.

\subsubsection{Cell Module D}

Cell module D is designed primarily for protein solution experiments, where the amount of sample is usually quite limited, thermal requirements are stringent, and only DLS90 and absorbance measurements are needed. These features are accounted for by using an optical cell with a reduced sample volume of 0.1 ml, less than one third of that of CM E, and by suppressing the unused CM windows. The thermal gradient and thermal stability thus obtained are $<$ 0.05°C/cm over the cell (in all directions), and $<$ 0.03°C over 10 hours, respectively. Furthermore, CM D allows for quenching temperature from 25°C to 10°C in less than 30 s, in order to trigger protein nucleation at a well-defined time. In addition to the gradient, stability and  $T$ ramp requirements, it is also required to increase/decrease the temperature in the range of 10 to 40°C. The stirring and temperature monitoring systems are  very similar to those of CM E.

\section{Scattering methods}
\label{sec:scattering_methods}

We briefly introduce the main features of the static and dynamic light scattering methods implemented in COLIS.

\subsection{Dynamic light scattering}
Dynamic light scattering is a well-established method to probe the dynamics of soft matter systems~\cite{berne76} on time scales ranging from 10 ns to about 100 s, and on length scales in the range 1 nm to about 1 µm, depending on the experimental geometry. In the COLIS implementation of DLS, the sample is illuminated by the optical laser beam with in-vacuo wave length $\lambda_0 = 532~\mathrm{nm}$ and light scattered at the three scattering angles mentioned in Sec.~\ref{sec:overview_COLIS} is collected by fiber optics and fed to fast, sensitive avalanche photodiode detectors. The information on the sample dynamics is encoded in the temporal fluctuations of the scattered intensity, $I(q,t)$, which are quantified by the time autocorrelation function
\begin{equation}
g_2(q,\tau)-1 = \frac{\left <I(q,t)I(q,t+\tau)\right>_t}{\left <I(q,t)\right>_t^2}-1 \,,
\label{eq:g2}
\end{equation}
where $q = 4\pi n \lambda_0^{-1} \sin(\theta/2)$ is the modulus of the scattering vector, $n$ the refractive index of the solvent, $\theta$ the scattering angle, and $<\dots>_t$ indicates a time average. Note that here we have assumed that the system is isotropic, such that $g_2-1$ does not depend on the orientation of the scattering vector. The intensity correlation function is related to the field correlation function $g_1$, also known as intermediate scattering function, by the Siegert relation:~\cite{berne76}
\begin{equation}
g_2(q,\tau)-1 = \alpha g_1^2(q,\tau)= \alpha \left [ \frac{\left < \sum_{j,k=1}^N\exp[-i\mb{q}\cdot(\mb{r}_j(0) - \mb{r}_k(\tau))]\right > }{\left < \sum_{j,k=1}^N\exp[-i\mb{q}\cdot(\mb{r}_j(0) - \mb{r}_k(0))]\right > } \right ]^2
\label{eq:Siegert}  \,,  
\end{equation}
where the coherence factor $\alpha$ is a setup-dependent constant~\cite{berne76}, close to 1 for COLIS. In the last equality of Eq.~\ref{eq:Siegert}, the term in the brackets is the (collective) intermediate scattering function $f(q,\tau)$, where we have assumed for simplicity that the scattering is due to $N$ identical particles, with time-dependent positions $\mb{r}_1,...,\mb{r}_N$, and that the system is ergodic and the dynamics stationary, such that the ensemble average $<\dots>$ in Eq.~\ref{eq:Siegert} is equivalent to the time average used to calculate $g_2-1$ .

COLIS allows for both polarized and depolarized DLS, thanks to polarizers placed in front of each DLS fiber optics. 
The conventional `polarized' DLS configuration, or VV configuration, is obtained by orienting the polarizers' transmission axis in the same direction as the polarization direction of the incoming beam. `Depolarized' dynamic light scattering (DDLS) corresponds to the so-called VH geometry, where the polarizers' axis is rotated by 90 degrees with respect to the incoming beam polarization. For optically isotropic particles, the VH scattered intensity is null. %
By contrast, the scattered intensity from either geometrically or optically anisotropic particles has both a VV and a VH component.  We recall here only a few results for DDLS from partially crystalline colloidal particles, of interest for the experiments discussed in Sec.~\ref{sec:Brownian_depolarized}, see Ref.~\cite{Piazza1994, Piazza1995} for a more detailed discussion. %
A DDLS experiment measures $g_2^{V H}(q,\tau)-1$, the autotcorrelation function of the depolarized scattered intensity. %
Assuming that the orientation of the optical axis of distinct particles is uncorrelated and that the particle orientation and translation are decoupled, one finds:
\begin{equation}
g_2^{V H}(q,\tau)-1 = \alpha \left [f_r(\tau)f_s(q,\tau) \right]^2 \,,
\label{eq:DDLS_general}
\end{equation}
where $f_s(q, t)$ is the translational self intermediate scattering function and $f_r(t)$ is its ($q$-independent) rotational counterpart. Thus, under the conditions mentioned above, DDLS differs from DLS in two important respects: it probes self rather than collective dynamics, and it is sensitive to both translational and rotational dynamics. %

\subsection{Space- and time-resolved dynamic light scattering: Photon Correlation Imaging  }
\label{sec:PCI}

\begin{figure}[h]
    \centering
    \includegraphics[width=1\linewidth]{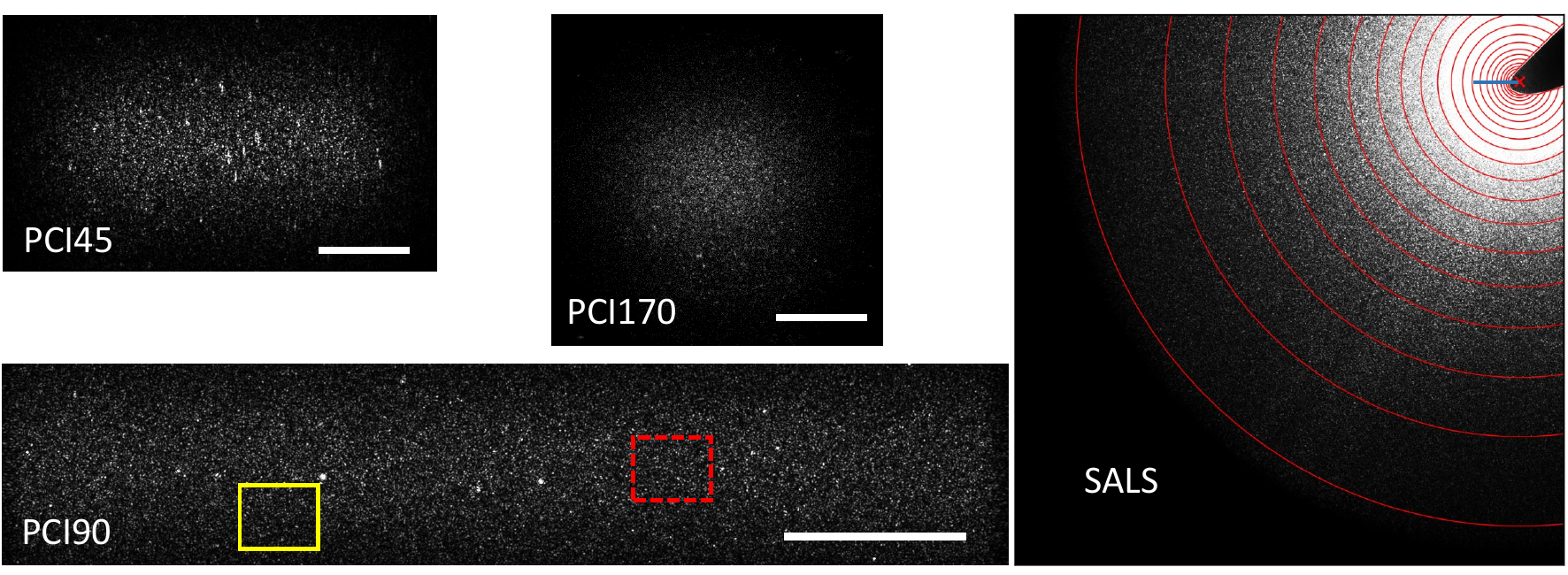}
    \caption{Typical speckle images images collected by the three COLIS PCI cameras and the SALS camera. The PCI images have been cropped to show only the scattering volume; the bar corresponds to 1 mm. Different ROIs in the PCI images, such as the two boxes shown for PCI90, correspond to distinct regions of the scattering volume, but to the same scattering angle. For the SALS camera, each annular ROI corresponds to light scattered by the whole illuminated sample within a narrow range of scattering angles. The beam block that stops the transmitted beam is visible in the top-right corner, the cross indicates the position of the transmitted beam, and the blue bar corresponds to a scattering angle $\theta = 1^{\circ}$.
    Pixels masked by the beam block are excluded from the annular ROIs. See Sec.~\ref{sec:SALS_general} for details on how the large variation of the scattered intensity with $\theta$ is dealt with, to avoid over- or under-exposing the image.}
    \label{fig:cameras}
\end{figure}

Soft solids such as gels and colloidal glasses often exhibit relaxation times as long as several hours and heterogeneous dynamics that evolve and fluctuate both in time and spatially. Conventional DLS can not cope with these features, because it relies on extensive time-averaging, which should typically last $10^3-10^4$ times longer than the longest relaxation time of the system~\cite{berne76}, and because the scattered intensity collected by the detector originates from the whole scattering volume. Photon Correlation Imaging~\cite{duri2009resolving} is a multispeckle technique~\cite{cipelletti_scattering_2016} that overcomes these limitations by using CMOS cameras as detectors and by combining scattering and imaging. In COLIS, three PCI cameras make an image of the scattering volume, using light scattered at the same angles as for conventional DLS. Due to the coherence of the illuminating beam, the small acceptance angle of the objective lenses used to form the image ($3.1^{\circ}$ for water-based samples, as set by a diaphragm placed in the focal plane of the objective lens), and the low magnification ($M = 1.38$%
), individual colloidal particles cannot be resolved in the PCI images. Rather, the images have a speckled appearance, as shown in Fig.~\ref{fig:cameras}. Distinct regions of interest (ROIs), such as the two boxes shown in the PCI 90 image, correspond to distinct regions within the scattering volume. Accordingly, the dynamics are quantified by a space- and time-resolved degree of correlation $c_I$:
\begin{equation}
\begin{aligned}
    \tilde{c}_I(\mb{r},q,t,\tau) = \frac{\left <I_p(q,t)I_p(q,t+\tau)\right>_{\mb{r}}}{\left <I_p(q,t)\right>_{\mb{r}}\left <I_p(q,t+\tau)\right>_{\mb{r}}}-1 \\
    {c}_I(\mb{r},q,t,\tau)  = \frac{\tilde{c}_I(\mb{r},q,t,\tau) }{\sqrt{\tilde{c}_I(\mb{r},q,t,0)\tilde{c}_I(\mb{r},q,t+\tau,0)}}\,,
\end{aligned}
\label{eq:g2:PCI}
\end{equation}
where $I_p$ is the intensity measured by the $p$-th pixel, $<\dots>_{\mb{r}}$ indicates an average over a ROI centered around position ${\mb{r}}$ in the sample, and where the normalization introduced in the second line results in $c_I(\tau=0) = 1$ and reduces the statistical noise associated to the finite number of pixels for $\tau > 0$~\cite{duri_timeresolvedcorrelation_2005}. For stationary, spatially uniform dynamics, averaging $c_I$ over $t$ and ${\mb{r}}$ yields the conventional $g_2-1$ function of
Eq.~\ref{eq:g2}. Note that CMOS cameras typically have non-negligible dark counts. Therefore, $I_p$ in Eq.~\ref{eq:g2:PCI} is obtained by subtracting a dark background from the CMOS signal, as detailed in~\cite{duri_timeresolvedcorrelation_2005}. All calculations required to obtain $c_I$ are performed on-board, using a dedicated, in-house software. The software allows also for correcting $c_I$ for any rigid drift of the speckle pattern within a ROI~\cite{cipelletti_simultaneous_2013a}, which may occur in gels or glasses due to the relaxation of internal stresses.

\subsection{Small-angle light scattering  }
\label{sec:SALS_general}
Small angle light scattering (SALS) is implemented in COLIS using a CMOS camera, allowing for both static and dynamic light scattering measurements. The optical layout is similar to that of Ref.~\cite{ferri_use_1997}: light scattered at small angles (typically, $0.25^{\circ} \le \theta \le 10^{\circ}$, corresponding to more than a decade and a half in scattering vector magnitude, $0.06~\mu\mathrm{m}^{-1} \le q \le 2.7~\mu\mathrm{m}^{-1}$) is collected by a Fourier lens; an objective lens images the focal plane of the Fourier lens onto the CMOS sensor. A typical SALS image is shown in Fig.~\ref{fig:cameras}. The transmitted beam is stopped by a beam block, placed in the focal plane of the Fourier lens; pixels at a distance $r_p$ from the transmitted beam position receive light exiting the cell at an angle $\arctan(r_p/f_{eff})$, where $f_{eff} = 38.18~\mathrm{mm}$ is the effective focal length of the SALS apparatus%
, which accounts for both the focal length of the Fourier lens and the magnification of the objective lens. 

As seen in Fig.~\ref{fig:cameras}, the SALS scattered intensity usually varies considerably with $\theta$, making it impossible to correctly capture $I(q)$ over the full range of SALS $q$ vectors using a single exposure time. To circumvent this problem, a burst of images at a predefined set of exposure times is acquired and a single composite image is reconstructed using, for each ROI, the data collected at the optimum exposure time. The ROIs have an annular shape, corresponding to a small range $\Delta \theta$ of scattering angles centered around a well defined $\theta$, with $\Delta \theta / \theta \lesssim 0.1$, and the optimum exposure time is chosen such that the ROI-averaged intensity is closest to, yet smaller than, an empirically determined threshold of 40 grey levels for 8-bit images. As for PCI, a dark background is subtracted systematically before further processing of the images.

The composite images are used for both dynamic and static SALS. For dynamic light scattering, the images are processed in the same way as PCI images, Sec.~\ref{sec:PCI}, the only difference being that all averages are performed on the annular ROIs, yielding one $c_I(q,t,\tau)$ dataset per ROI. Note that there is no $\mb{r}$ dependence in the SALS $c_I$: SALS data are collected in the far field geometry and thus lack spatial resolution, in the sense that each pixel receives light from the whole scattering volume. Static SALS is performed by calculating the intensity of the scattered light averaged over all pixels belonging to a given ROI and by applying several normalization factors:
\begin{equation}
I(q,t) = \frac{C(q)E_{ref}}{PDT(t)E_{opt}(q,t)}\left<I_p(t)\right>_q \,,
\label{eq:SALS}    
\end{equation}
where $<\dots>_{q}$ indicates the average over all pixels belonging to the annular ROI associated to a scattering vector of magnitude $q$, $E_{opt}(q,t)$ is the time- and $q$-dependent chosen optimum exposure time and $E_{ref}$ is an arbitrary reference exposure time, set to 1 ms in COLIS. The normalization factor $C(q)$ accounts for the small variation of the solid angle associated with each pixel (less than 10\% over the angular range of COLIS, see Ref.~\cite{tamborini_multiangle_2012} for details), and $PDT(t)$ is the intensity of the incident beam at the time of the acquisition, obtained from the monitor photodiode mentioned in Sec.~\ref{sec:overview_COLIS}. Although no absolute intensity calibration is available in COLIS, the normalization factors included in Eq.~\ref{eq:SALS} allow for comparing the relative intensity between different samples.

Small angle scattering is notoriously challenging because any imperfection in the optical elements (lenses, cell walls, additional windows introduced to meet safety rules imposing multiple containment levels, etc.) scatters light in the SALS angular range. Correction schemes for this so-called optical background are possible for both dynamic~\cite{luca_cipelletti_ultralow-angle_1999} and static measurements~\cite{tamborini_multiangle_2012}, provided that one can measure the scattering pattern with the cell filled with only the solvent, using the very same cell that will be later loaded with the sample. Because COLIS is designed to operate with various cells, some of which will not be available prior to the upload of the setup onboard the ISS, these correction schemes will not be always applicable. Examples of SALS data with and without optical background correction will be presented in Sec.~\ref{sec:SALS_results}.

\section{Results} \label{sec:results}

We illustrate most of the capabilities of COLIS, starting from validation tests on model colloidal particles (Secs.~\ref{sec:Brownian}-\ref{sec:SALS_results}) and then moving to more complex protein, gel and glassy systems (Secs.~\ref{sec:NIR_tests}-\ref{sec:glass}).

\subsection{Brownian dynamics of model colloidal particles}\label{sec:Brownian}

\subsubsection{Polarized DLS  }\label{sec:Brownian_DLS}

\begin{figure}
\centering
\includegraphics[width=0.95\textwidth]{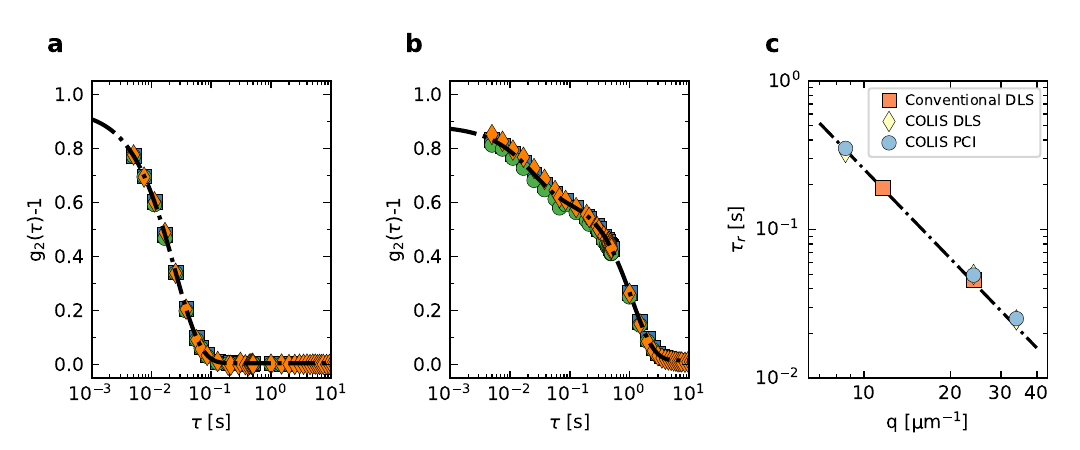}
\caption{%
Dynamics of Brownian particles measured by polarized light scattering. a) intensity correlation function measured by the PCI90 camera. Symbols: data from three independent runs; line: fit with Eq.~\ref{eq:g2Brownian} b): same as in a), for $\theta = 170.8$ deg. The line is a double exponential fit, Eq.~\ref{eq:g2BrownianTwoContributions}, see text for details. c): fitted relaxation time as obtained by COLIS (PCI and DLS detectors) and by a conventional setup, as a function of scattering vector $q$. The line is a fit to the data using the power law $\tau_r \propto q^{-2}$ predicted for Brownian diffusion. Error bars, as obtained from the fit, are smaller than the symbol size. %
}
\label{fig:Brownian}
\end{figure}

We test VV, or polarized, dynamic light scattering at the three wide angles covered by COLIS by measuring the Brownian dynamics of a diluted suspension of polystyrene (PS) particles, using both the avalanche photodiode detectors and hardware correlators (DLS, hereafter) and the PCI cameras. For the latter, we use the variable delay acquisition scheme of Ref.~\cite{philippe_efficient_2016}, which allows probing $\tau$ delays as small as 2 ms while keeping low the average image acquisition rate (here, 2 Hz). All measurements are performed simultaneously. Figure~\ref{fig:Brownian}a) shows data for the PCI 90 deg camera, for three successive runs, showing excellent repeatability. The intensity correlation function, obtained by averaging $c_I$ over the full scattering volume and over time, is fitted by (line in Fig.~\ref{fig:Brownian}a):
\begin{equation}
g_2(q,\tau)-1 = A\exp[-2\tau/\tau_r] + B \,,
\label{eq:g2Brownian}
\end{equation}
where for Brownian particles the relaxation time reads $\tau_r = (Dq^2)^{-1}$~\cite{berne76}, with $D$ the (translational) diffusion coefficient given by the Stokes-Einstein relation, $D = k_BT/(6\pi\eta R)$, with $R$ the particle radius $R$, $\eta$ the solvent viscosity, $T$ the absolute temperature and $k_B$ Boltzmann's constant. $A \lesssim 1$ is a constant that depends on the speckle-to-pixel size ratio~\cite{goodman_speckle_2007} and $B \approx 0$ is the baseline~\cite{duri_timeresolvedcorrelation_2005}. Note that, since CMOS cameras are slower detectors compared to the avalanche photodiodes used for DLS, the smallest delay time $\tau$ accessible by PCI is here 5 ms, still sufficient to correctly capture the particles' dynamics. 

A similar exponential relaxation is seen for the DLS data at $\theta = 90$ deg and for both the TRC and DLS data at nominal 45 deg ($\theta = 29.32~\mathrm{deg}$, see Fig. S4 of SI). By contrast, the correlation function measured at $\theta  = 170.8$ deg %
exhibits a surprising two step decay, Fig.~\ref{fig:Brownian}b. This behavior is reminiscent of that reported in Ref.~\cite{pommella_coupling_2019}. It is due to the reflection of part of the incident beam at the cell exit wall. This reflected light propagates backward in the scattering cell, along the same axis as for the incident beam, but in the opposite direction. As a result, the detector at $\theta = 170.8$ deg collects both backscattered light and light scattered in the forward direction by particles illuminated by the counter-propagating reflected beam. The intensity correlation function can then be modelled by the squared sum of two exponential relaxations, corresponding to the two contributions (line in Fig.~\ref{fig:Brownian}b); further assuming Brownian dynamics, leads to
\begin{equation}
g_2(q,\tau)-1 = \left [ A_f\exp(-Dq_f^2 \tau) + A_b\exp(-Dq_b^2 \tau) \right ]^2 + B \,,
\label{eq:g2BrownianTwoContributions}
\end{equation}
where the indexes $f$ and $b$ refer to forward- and back-scattering, respectively.

Figure~\ref{fig:Brownian}c shows the $q$ dependence of the relaxation time $\tau_r = (Dq^2)^{-1}$ obtained from the fits of $g_2-1$ (Eq.~\ref{eq:g2Brownian} for data at $\theta =  29.32$ and 90 deg, backscattering term of Eq.~\ref{eq:g2BrownianTwoContributions} for the data at $\theta = 170.8$). Additional measurement performed on the same sample using a conventional, commercial setup (Brookhaven BI-9000AT) are also shown (red squares). All data fall onto the same straight line on a double logarithmic plot, corresponding to the $q^{-2}$  scaling expected for Brownian dynamics. A power law fit to the data, $\tau_r = D^{-1} q^{-2}$, yields $D = 0.0390 \pm 0.0006~\mu\mathrm{m}^2\mathrm{s}^{-1}$, in good agreement with $D= 0.042~\mu\mathrm{m}^2\mathrm{s}^{-1}$ as calculated using the nominal particle size and nominal solvent viscosity.

\subsubsection{Depolarized DLS   }\label{sec:Brownian_depolarized}

\begin{figure}
\centering
\includegraphics[width=0.95\textwidth]{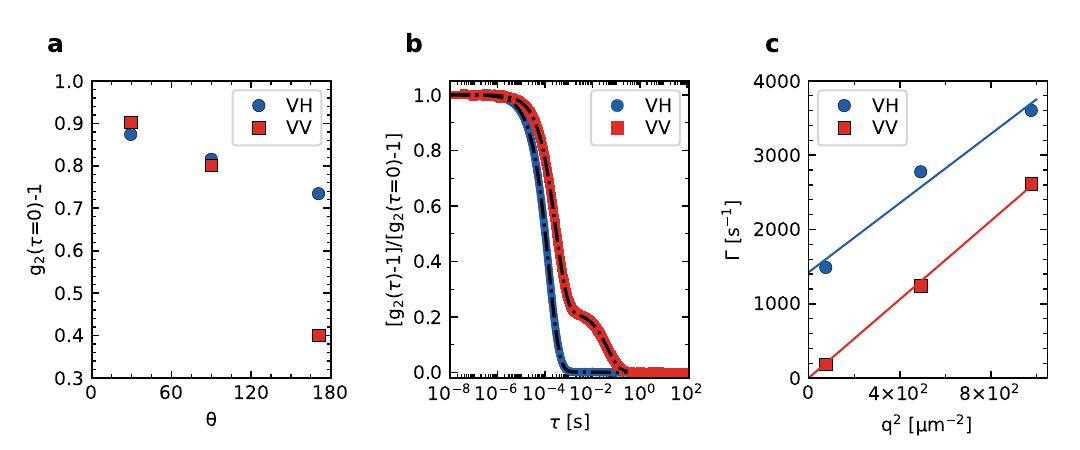}
\caption{Translational and rotational dynamics of MFA colloids probed by DLS and DDLS. a) $g_2(\tau\rightarrow 0)-1$ \textit{vs} scattering angle for polarized (VV, red squares) and depolarized (VH, blue circles) conditions. Error bars, as obtained from the standard deviation over four independent repetitions, are smaller than the symbol size. b): VV and VH intensity correlation functions measured at $\theta = 170.8^{\circ}$ with COLIS DLS170. Continuous lines are fits as discussed in the main text. c) $q$ dependence of the decay rates obtained from the fits to $g_2-1$, for VV and VH conditions. Error bars, as obtained as for a), are smaller than the symbol size. %
}
\label{fig:DepDLS_bis}
\end{figure}

We show data obtained for a diluted suspension in water of Brownian, non-interacting and nearly monodisperse particles made of Hyflon MFA, a copolymer of tetrafluoroethylene and perfluoromethylvinylether. Under these conditions, the expected functional form for the DLS $g_2-1$ is given by Eq.~\ref{eq:g2Brownian}, while for DDLS Eq.~\ref{eq:DDLS_general} reduces to
\begin{equation}
g_2^{V H}(q,\tau)-1 = A\exp\left [-2(q^2D+6D_r)\tau\right] + B,\,
\label{eq:DDLS_Brownian}
\end{equation}
where $D$ and $D_r = k_BT/(8\pi \eta R^3)$ are the infinite-dilution translational and rotational diffusion coefficients of a single particle, respectively, and $A \lesssim 1 $ and $B \approx 0$ are as in Eq.~\ref{eq:g2Brownian}. Note that the relaxation of $g_2^{V H}-1$ is exponential, as for DLS, but with a modified decay rate $\Gamma \equiv \tau_r^{-1} = q^2D+6D_r$.

We run simultaneous DLS measurements at $T=25~^{\circ}$C  using the DLS45, DLS90, and DLS170 COLIS lines, which correspond to $\theta = 29.32^\circ,~90.0^\circ$, and $170.8^\circ$, and $q = 8.7~\si{\micro\meter}^{-1},~22.2~\si{\micro\meter}^{-1}$ and $31.3~\si{\micro\meter}^{-1}$, respectively. %
Measurements are run in both the VV and the VH configuration. 
Figure~\ref{fig:DepDLS_bis}(a) shows the $\theta$ dependence of the intercept of the correlation functions, i.e., their $\tau \rightarrow 0$ value. For the VV component, the intercept drops sharply for backscattering, most likely due to the contribution of stray light stemming from the back reflection of the transmitted beam on the cell exit wall, as discussed in Sec.~\ref{sec:Brownian_DLS}. By contrast, the intercept of the depolarized correlation function exhibits only a mild $\theta$ dependence, because the crossed polarizer blocks the spurious back-reflected light, responsible for the lower VV intercept. 

The presence of a spurious contribution in the DLS170 data, due to unwanted back-reflected light, is confirmed by inspecting the shape of $g_2-1$. As shown in Fig.~\ref{fig:DepDLS_bis}b, $g_2^{VV}(\tau)-1$ exhibits two distinct decay times, as already observed in the polarized DLS data of Fig.~\ref{fig:Brownian}b. By contrast, the $g_2^{VH}(\tau)-1$ exhibits a nearly exponential decay, as predicted by Eq.~\ref{eq:DDLS_Brownian}. A careful inspection of the correlation function reveals very small deviations from a simple exponential relaxation, which we attribute to a small fraction of the VV light passing through the polarizer, due to its imperfect alignment and to the small birefringence of the optical windows and lenses. Accordingly, we fit the DDLS correlation function measured by DLS170 with the following expression:
\begin{equation}
g_2^{V H}(q,\tau)-1 =(1-a)\exp\left [-2(q^2D+6D_r)\tau\right] + a\left [g_2^{(\mathrm{emp})}(q,\tau)-1\right],\,
\label{eq:DDLS170_Brownian}
\end{equation}
where the first term on the right-hand side accounts for the expected DDLS contribution, while the second one is an empirical fit to the VV contribution (red squares in Fig.~\ref{fig:DepDLS_bis}b, using Eq.~\ref{eq:g2BrownianTwoContributions}. The coefficient $a$ accounts for the relative weight of the VV and VH components, yielding an effective extinction ratio of the polarization optics of COLIS of 0.02. 
Figure~\ref{fig:DepDLS_bis}(c) shows the decay rate $\Gamma = \tau_r^{-1}$ (symbols), obtained from the fits to the VV and VH correlation functions. Fitting the VV relaxation rate to $\Gamma = Dq^2$ (red line in Fig.~\ref{fig:DepDLS_bis}c) and using the Stokes-Einstein relation with $\eta = 0.89~\si{mPa ~s}^{-1}$ yields $R = 92 \pm 2$~nm.%
The VH relaxation rate is consistent with the expected behavior $\Gamma=6D_R+Dq^2$ (blue line in Fig.~\ref{fig:DepDLS_bis}c). The particle radius may be estimated from $D$, yielding $R=105 \pm 10$ nm, or using the ratio between the translational and rotational diffusion coefficients, yielding $R=\sqrt{3D/4D_R} = 85 \pm 10$~nm. %
By averaging the $R$ values issued from the three methods we obtain $R = 93 \pm 5$~nm, in excellent agreement with the expected value, $R = 90 \pm 2$~nm.

\subsection{Small-angle static and dynamic light scattering  }
\label{sec:SALS_results}

\begin{figure}
    \centering
    \includegraphics[width=1\linewidth]{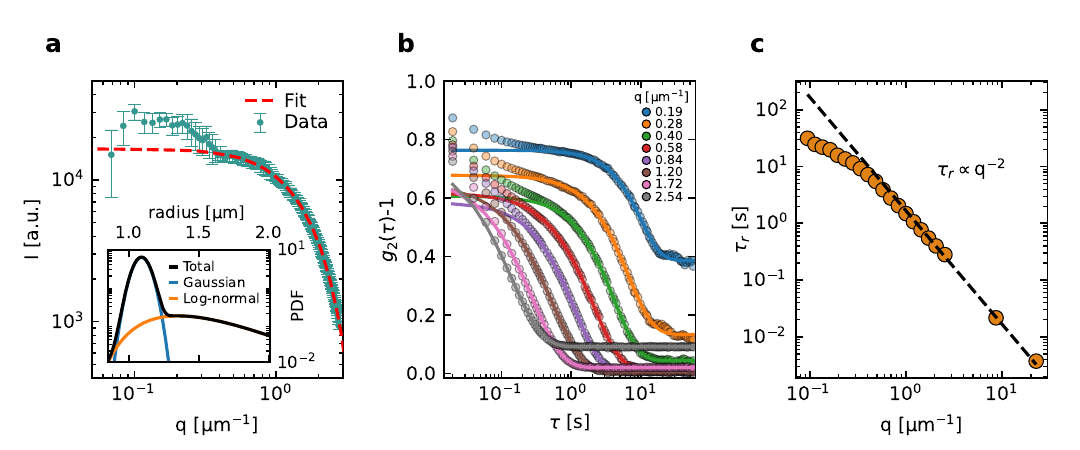}
    \caption{%
    SALS static and dynamic light scattering on diluted colloidal suspensions. (a): static scattered intensity vs $q$ after optical background subtraction, for a diluted suspension of SiO$_2$ colloids. Error bars are the standard deviation over a time series of scattering images. (b) Representative $g_2-1$ functions measured with the SALS camera for a a diluted suspension of PS particles, with no optical background subtraction. (c) $q$ dependence of the relaxation time of the correlation functions shown in (b), with additional data points obtained a larger angles, using the DLS45 and DLS90 COLIS measurement lines. Error-bars are calculated using the covariance matrix from the fits and are smaller than the symbols. }
    \label{fig:SALS}
\end{figure}

Figure~\ref{fig:SALS} shows static and dynamic light scattering data (SALS-SLS and SALS-DLS, respectively) obtained for diluted suspensions of colloidal particles using the SALS camera of COLIS. 
The SALS-SLS data for SiO$_2$ particles shown in Fig.~\ref{fig:SALS}a) were corrected for the optical background, taken with the cell filled with the solvent alone, see Sec.~\ref{sec:SALS_general}. Data for $q \ge 0.4~\mu \mathrm{m}^{-1}$ show the characteristic decrease of $I(q)$ expected for spherical particles. At smaller $q$, the scattered intensity becomes more erratic and tends to be higher than expected. This behavior is typical of scattering data collected at very small $q$, where the optical background is comparable to or even larger than the signal due to the particles, making it difficult to fully correct for stray light. Indeed, at the smallest $q$ the optical background for the data of Fig.~\ref{fig:SALS}a was more than five times larger than the particles' contribution. The inset shows the particle size distribution obtained using Mie theory and modelling the suspension as a mixture of quasi-monodisperse spherical particles (Gaussian distribution with an adjustable mean value and fixed relative standard deviation of 5\%, as provided by the manufacturer) and a log-normally distributed population of larger effective spheres, accounting for any aggregates that may be present. The fit is performed for $q \ge 0.4~\mu \mathrm{m}^{-1}$; as shown by the dotted line in Fig.~\ref{fig:SALS}a, it reproduces very well $I(q)$. The Gaussian distribution accounts for 85.6\% of the scattered intensity and corresponds to an average particle radius of $1.09~\mu \mathrm{m}$, in very good agreement with the nominal one ($1.005~\mu \mathrm{m}$). The log-normal distribution corresponds to a small amount of aggregates with effective radius $1.54~\mu \mathrm{m}$ and moderate polydispersity (relative standard deviation: 25\%).%

SALS-DLS data for Brownian PS particles were collected without taking any optical background, to test the conditions of some of the space experiments, see Sec.~\ref{sec:SALS_general}. Representative $g_2-1$ functions measured simultaneously at various $q$ vectors are shown in Fig.~\ref{fig:SALS}b. As $q$ decreases, the decay time increases, as expected, and the base line increasingly departs from zero, due to the contribution of the static stray light~\cite{luca_cipelletti_ultralow-angle_1999}. The fast relaxation mode seen for $\tau \lesssim 0.1$~s is possibly due to aggregates or impurities that sediment through the beam. We fit the final decay of $g_2-1$ using Eq.~\ref{eq:g2Brownian} with $A$, $B$, and the relaxation time $\tau_r$ as the fitting parameter. The fitted relaxation time is shown in Fig.~\ref{fig:SALS}c as a function of $q$, for the SALS data and for additional DLS measurements performed on the same sample using the COLIS DLS90 and DLS45 detection lines. The SALS relaxation time agrees well with the expected behavior for $q \gtrsim 0.5~\mu \mathrm{m}^{-1}$: in this regime, the SALS and DLS relaxation times are very well fitted by $\tau_r = (Dq^2)^{-1}$ (dotted line), with $D = 0.60 \pm 0.01~\mu \mathrm{m}^{2}\mathrm{s}^{-1}$%
, in excellent agreement with $D = 0.59~\si{\micro m}^2\si{s} ^{-1}$ calculated from the nominal particle radius and solvent viscosity. At lower $q$, the relaxation time of $g_2-1$ is faster than expected, possibly due to residual sedimentation or convection.%

\subsection{Space-resolved measures of NIR-induced heating  }
\label{sec:NIR_tests}
\begin{figure}
\centering
\includegraphics[width=0.75\textwidth]{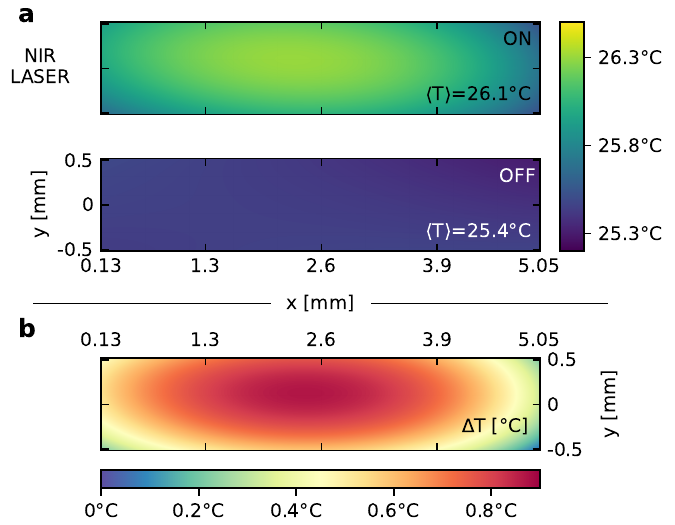}
\caption{%
Map of the local temperature while illuminating the scattering volume with the NIR laser. The cell temperature is controlled to $T= 25.5~^{\circ}$C with the Peltier elements and the NIR laser is switched on during the measurement, with a power of $60$~mW. \textbf{(a)}: Local temperature during the 'ON' and 'OFF' status, respectively. The maps have been calculated averaging 180 and 20 frames for the OFF and ON status, respectively, see details in the Supplementary Information. Note the hotter region in the middle of the sample, due to the boundary conditions imposed by the Peltier elements. \textbf{(b)}: Local change of $T$, calculated as the difference of the maps shown in panel \textbf{(a)}. 
}
\label{fig:Tmap}
\end{figure}

As a way to demonstrate both the space resolution capabilities of PCI and the local heating of the scattering volume through the NIR laser, we build a map of the local temperature upon illuminating the scattering volume with the NIR beam. We use a colloidal glass comprising pNIPAM microgel particles, a sample with two unique properties. First, convection is suppressed, because the sample is glassy, thus avoiding mixing on macroscopic length scales when the NIR laser is turned on. Second, the scattered intensity changes upon heating, because the microgels shrink~\cite{brugnoni_synthesis_2019}, resulting in changes of both the form factor of the particles and the structure factor of the suspension; thus, $I = I(q,T)$. We calibrate the temperature dependence of $I$ by acquiring data with the PCI90 camera, with no NIR beam and using the cell temperature control to vary systematically $T$ from $T_{min}=25\,^\circ$C up to $T_{max}=30\,^\circ$C, acquiring 16 different temperature points (more details in the Supplemental Material). For each camera pixel $p$, a $T$-dependent intensity value is obtained, allowing for building a set of $I_p(T)$ calibration curves, see the SI for full details of the procedure. Therefore, we are able to convert the intensity values with the NIR beam on to a map of the local temperature. An example is shown in Fig.~\ref{fig:Tmap}, which displays the $T$ map obtained with the cell temperature set to $T= 27~^{\circ}$C, both before and 300 s after switching the NIR beam on, when the $T$ field has reached a steady state. Note that the temperature increase is as high as $1$°C, with the NIR laser running at just 20\% of its maximum power. A higher NIR beam intensity induces a slow drift of the overall cell temperature, due to limitations in the power of the Peltier elements.  %

\subsection{Using COLIS to study protein nucleation  }\label{sec:protein}
One of the planned COLIS experiments onboard the ISS is the investigation of protein nucleation, the first step of crystallization, which is a process of critical importance in the production of pharmaceuticals and in human health, as well as the object of a great deal of fundamental research. Protein nucleation frequently involves multiple steps and is affected by the presence of poorly understood nucleation precursors~\cite{precursors}. Our experiment will make use of the unique capabilities of COLIS to very precisely control temperature and temperature changes, while monitoring populations of clusters from molecular monomers to large precursor clusters. 

\begin{figure}
    \centering
    \includegraphics[width=1\linewidth]{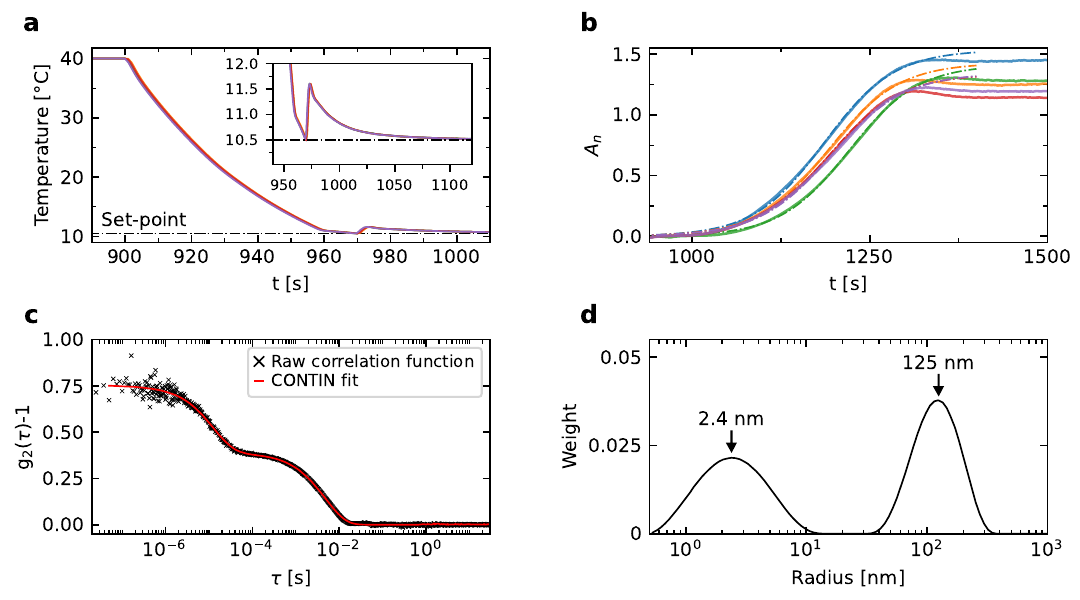}
    \caption{COLIS results from protein nucleation studies. (a): measured temperature as a function of time. The results of five independent runs are shown. The horizontal line at the bottom is the target final temperature. The inset shows that the undershoot is minimal. (b): normalized absorbance %
    from five repetitions of the cycle of cooling to 10.5°C followed by heating to 40°C. Dashed lines are fits of each curve to a sigmoid (only using data up to the maximum of each curve). (c): experimental and fitted DLS correlation functions. (d) distribution of cluster sizes obtained from the CONTIN fit shown in (c). The smaller peak is centered at 2.4 nm which is typical of a lysozyme molecule. The second peak, at 125 nm, is typical of nucleation precursors.}
    \label{Fig_Protein_1}
\end{figure}

Our experiments involve heating a dilute solution of proteins to $40$°C so as to dissolve any crystals present, and then to reduce the temperature to a set-point in the range of $10-15$°C as quickly as possible with minimal undershooting. The latter point is important because even if the temperature returns to the set-point, any sustained undershoot will trigger nucleation and crystal growth faster than at the set-point. Figure~\ref{Fig_Protein_1}(a) shows the results of five repetitions of the cooling protocol developed in the course of ground tests. The main panel and the inset clearly demonstrate the high degree of reproducibility of the cooling curves, the rapid rate of temperature decrease (approx. 0.5°C/sec) and the very low undershoot (less than 0.015°C).   

The process of nucleation - that is, the formation, via thermal fluctuations, of crystalline clusters large enough to be stable and grow - will be monitored by means of DLS and absorbance. Absorbance measurements are made by measuring the intensity of the laser beam, before and after passing through the sample, $I_0$ and $I_t$, respectively. %
The absorbance is defined as $A = \log_{10} (I_0/I_t)$; in Fig.\ref{Fig_Protein_1}b we show $A_n$, the absorbance normalized by its value at time $t =0$, when the proteins are fully dispersed in the solution, during one cycle of temperature reduction from 40°C to a set-point of 10.5°C, from our ground experiments on the protein lysozyme. The absorbance initially follows a sigmoidal curve, as expected, from which the induction time for nucleation can be extracted by fitting the data. (Only data up to the maximum of each curve is used in fits since, at later times, the curves are no longer sigmoids due to larger crystals falling out of the field of view in terrestrial gravity.) The fits show  induction beginning at $285 \pm 6$~s after start of the temperature decrease, thus illustrating the accuracy with which the induction time can be measured from the experimental data.

The second means of monitoring the process is DLS. Figure \ref{Fig_Protein_1}(c) shows an example of a correlation function $g_2-1$ measured using DLS90 during a period of 30 seconds, at the end of the melting cycle at $40$°C. Note the double shoulder in $g_2-1$, which indicates at least two populations of different-sized scatterers. The  data has been processed using the CONTIN~\cite{contin} algorithm from the Jscatter package~\cite{Jscatter} and the resulting fitted signal is shown together with the experimental $g_2-1$, demonstrating good agreement between the two. From this, we have information about the size and size distribution of any clusters present in the solution. For the example shown, the analysis indicates two populations of objects (\ref{Fig_Protein_1}(d)), one with a hydrodynamic radius of 2.4 nm, the typical size of a lysozyme molecule, and one with a mean radius of 125nm, which is typical of the aforementioned nucleation precursors~\cite{precursors}.

\subsection{Gelation and collapse of a physical gel  }\label{sec:gel}

\begin{figure}
    \centering
    \includegraphics[width=1\linewidth]{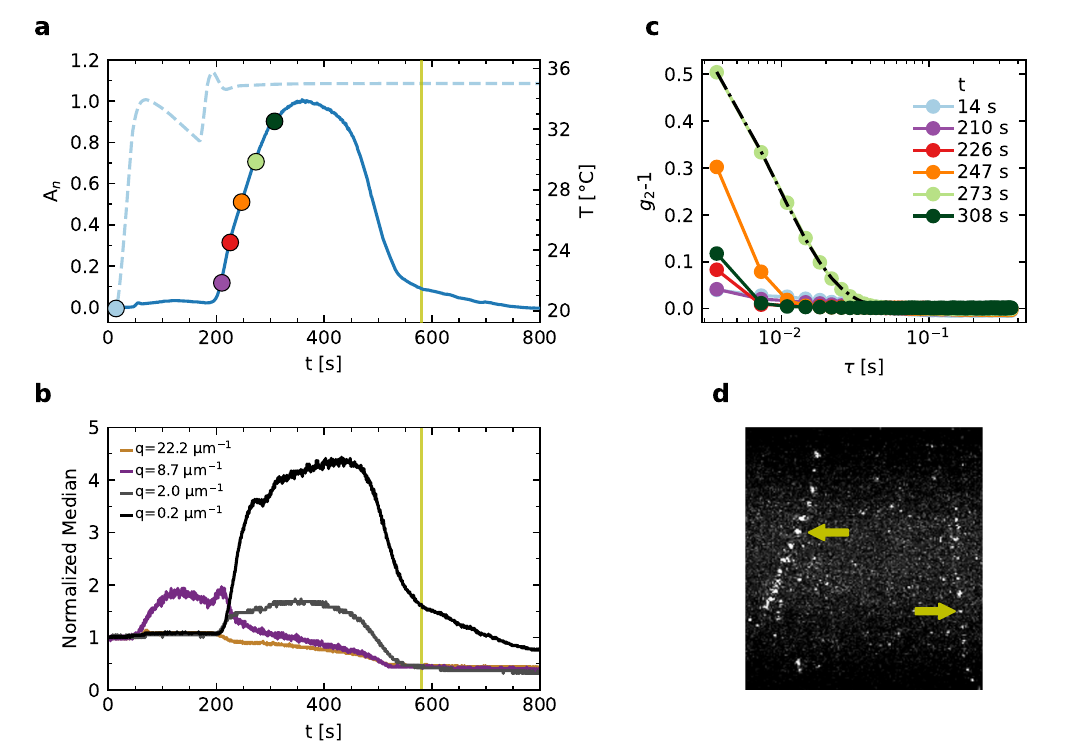}
    \caption{Gravity-suppressed gelation of attractive colloids. (a): Sample temperature (dashed line) and normalized absorbance (solid line) \textit{vs} time. 
    (b): Median value of the scattered intensity measured by the three PCI cameras and at one selected angle of the SALS camera, corresponding to the $q$ values shown in the label. (c): Intensity correlation functions measured by PCI90 at various times $t$, shown as circles with the same color code in (a). (d) Maximum intensity projection of a burst of 15 images taken by the PCI90 camera at the time shown by the vertical lines in (a) and (b). The arrows show two trajectories of bright spots that correspond to settling colloidal aggregates.}
    \label{Fig_gel}
\end{figure}

In space, COLIS will be used to investigate the slow restructuring of gels formed by attractive colloidal particles. In terrestrial gravity, the gels are unstable: they collapse or even fail to fully form due to their own weight. Here, we show how data collected with various lines of COLIS may be combined to gain a thorough picture of the impact of gravity on colloidal aggregation and gelation. In our system, short-range attractive interparticle interactions are due to depletion forces~\cite{AsakuraInteractionBodiesImmersed1954} arising from the osmotic pressure exerted by micelles of Triton TX100, a surfactant molecule. As mentioned in Sec.~\ref{sec:materials}, the strength of these forces may be tuned by varying $T$, because Triton X100 exhibits an inverted consolution gap with water: due to the growth of pre-critical fluctuations of the micelle concentration, an increase in the sample temperature is associated with a strengthening of depletion effects~\cite{Degiorgio1996, Buzzaccaro2010, Piazza2011}.  
We choose a sample composition such that the strength of attractive interactions steeply increases on approaching $T = T_{gel} \approx 33.5^{\circ}$C, effectively triggering colloidal aggregation.
Figure~\ref{Fig_gel}(a) shows the time evolution of $T$ during a fast upward $T$ jump. Note that, unlike the case of proteins discussed in Sec.~\ref{sec:protein}, here it is not critical to avoid under- and overshoots of $T$; accordingly, the temperature control parameters were optimized for achieving a fast up-ramp, resulting in some $T$ oscillations before reaching the target temperature of $35^{\circ}$C, $1.5^{\circ}$C above $T_{gel}$. The resulting absorbance normalized to its peak value,$A_n$,  is shown in Fig.~\ref{Fig_gel}a as a solid line. $A_n$ is low for $T< T_{gel}$, because the suspension is fully dispersed. As $T$ exceeds $T_{gel}$, the absorbance strongly increases, due to the formation of colloidal aggregates driven by strong depletion interactions. Note however that $A_n$ eventually decreases, for $t \gtrsim 350~$s, which is incompatible with the formation of a stable gel structure at constant $T > T_{gel}$, a first hint of gravity-induced gel disruption.

This is confirmed by the time evolution of the median scattered intensity\footnote{Here, we use the median instead of the more common arithmetic mean because the aggregation process results in a strong increase of the scattered light. As a consequence, a significant fraction of the camera pixels are saturated and the median is a more robust statistical estimator of the average scattered intensity.} at representative $q$ vectors, obtained from the COLIS cameras and shown in Fig.~\ref{Fig_gel}b.
The aggregation process is characterized by a strong increase in the scattered light at all angles. Previous works shows that, for colloidal aggregation resulting in fractal aggregates at fixed particle number density, the scattered intensity at any given $q$ initially increases and then plateaus to a fixed value, see, e.g., Ref.~\cite{cipelletti2000universal}. Furthermore, the smaller the $q$ vector, the longer it takes for $I(q)$ to reach its plateau value. Both features are seen in the early stages of the time evolution shown in Fig.~\ref{Fig_gel}b. However, the scattered intensity eventually drops at all probed $q$, implying a decrease of the total number of particles in the scattering volume, presumably due to sedimentation.

To gain further insight in the aggregation process, we collect several bursts of 1000 images each, acquired with the PCI90 camera run at 275 fps. A first burst of images is taken before increasing $T$, when the colloids are fully dispersed. Additional bursts are triggered automatically when the absorbance level first exceeds a threshold level, set to  $A_n = 0.1, 0.3, 0.5, 0.7, 0.9$, respectively (circles in Fig.~\ref{Fig_gel}a).
The intensity correlation functions $g_2-1$ calculated for each series of images are shown in Fig.~\ref{Fig_gel}c. Note that here the multispeckle approach afforded by PCI is essential: replacing the extensive time average required by conventional DLS with the pixel-averaging of Eqs.~\ref{eq:g2:PCI} allows for following the rapidly-evolving dynamics following the onset of aggregation.

At the beginning of the aggregation process ($t \le 210~\si{s}$) the dynamics of the sample are very rapid and indistinguishable from that of the un-aggregated sample: at the minimum accessible delay, $\tau=3.6~\si{\m s}$, $g_2-1$ is close to zero. Starting from $t=226~\si{s}$ ($A_n = 0.3$) the dynamics steadily slow down until $t=273~\si{s}$, then accelerate again, see the data for $t=308$~s. The slowing down of the dynamics can be attributed to the formation of clusters of increasing size. Assuming for simplicity that the decay of $g_2-1$ is due to the translational diffusion of non-interacting clusters, the typical cluster size measured at $t=273$~s by fitting the data with Eq.~\ref{eq:g2Brownian} (dash-dotted line) is $R= 2.42~\si{\micro \m}$, more than a factor of 25 larger than the particle size. The acceleration of the dynamics at later time is consistent with the scenario of falling clusters suggested by the non-monotonic behavior of $A_n$ and of the median intensity, Figs.~\ref{Fig_gel}(a) and (b), respectively. Visual inspection of the PCI90 images confirms this scenario. Figure~\ref{Fig_gel}(d) shows an image obtained applying a Maximum Intensity Projection filter (MIP) to 15 consecutive frames taken at $t=580$~s (vertical line in Figs.~\ref{Fig_gel}(a), (b)). Each pixel of the output MIP image is set to the maximum intensity recorded over the series of images. This allows for following the trajectory of falling clusters, which appear as bright specs in the PCI90 images, see arrows in 
Fig.~\ref{Fig_gel}d. Note that the trajectories deviate from the vertical direction due to the mixing flow in the cell induced by the falling clusters.

We emphasize that here the imaging collection optics of the PCI90 camera is essential in order to directly observe the falling clusters: had a conventional far field configuration be chosen, the speckle images would have been uniform, with no hints of the individual contribution of localized scatterers. The PCI90 images allow for rationalizing the results obtained with all other measuring lines: as the largest clusters rapidly settle to the bottom of the cell, the scattering volume is left with smaller aggregates and individual particles, leading to the decrease of $A_n$, of the median intensity and of the relaxation time of $g_2-1$ (Figs.~\ref{Fig_gel}(a), (b) and (c), respectively).

\subsection{Spontaneous and driven dynamics of a soft colloidal glass  }\label{sec:glass}
We use a colloidal glass composed of thermo-sensitive pNIPAM soft microgels to demonstrate the space- and time- resolved capabilities of the COLIS setup, as well as to explore the effect of a thermal perturbation on these out-of-equilibrium samples. In colloidal systems, the glass transition is driven by particle crowding: the control parameter is the volume fraction $\varphi$ occupied by the particles~\cite{Hunterphysicscolloidalglass2012}. Thermosensitive colloids allow for conveniently varying $\varphi$ by shrinking or swelling the particles, while working with a single sample at a fixed number density of colloids. This unique feature has made thermosensitive particles very popular; in particular, pNIPAM thermosensitive microgels are now regarded as a model system for investigating phase transitions of various kinds~\cite{yunker2014physics}, including the glass transition of soft particles~\cite{mattsson_soft_2009,philippe2018glass}.

\subsubsection{Spontaneous dynamics: aging and temporal heterogeneity  }
\label{sec:glass_spontaneous}
%

\begin{figure}
\centering
\includegraphics[width=1\textwidth]{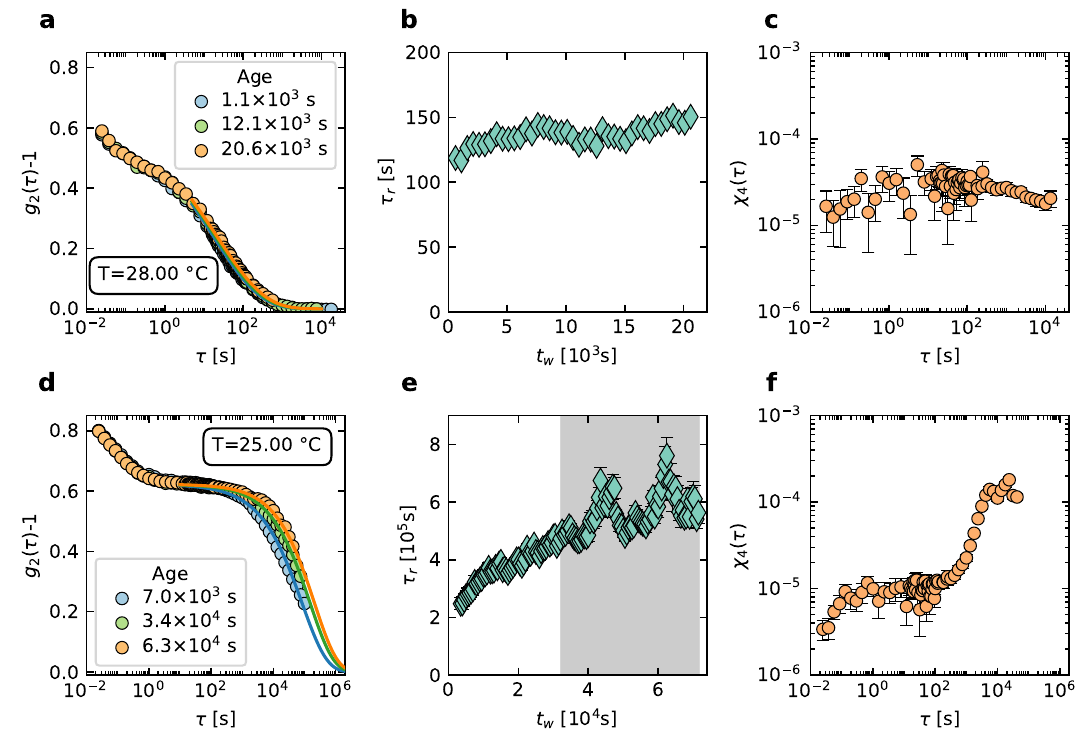}
\caption{Dynamics for a pNIPAM colloidal suspension in the supercooled ($T=28.00\,^\circ$C, top row) and glassy ($T=25.00\,^\circ$C, bottom row) states, respectively. Data obtained with  PCI90, $q=22.2 \,\si{\micro\meter}^{-1}$. a): intensity correlation functions (symbols), averaged over time windows of $500\,$s, for various sample ages $t_w$. Lines are fits with Eq.~\ref{Eq:KWW}. b): Relaxation time \textit{vs} $t_w$, showing only marginal aging. c): dynamical susceptibility quantify temporal heterogeneity.
d): same as a), but for $T=25.00\,^\circ$C. e) Relaxation time \textit{vs} $t_w$, note the strong fluctuations for $t_w \ge 3.2\times 10^{4}$~s (grey shaded region). f): dynamical susceptibility during the age interval marked in grey in e), hinting at the characteristic peak observed for glassy, heterogeneous dynamics. Panels b), c), e), and f): error-bars (1 standard deviation are calculated using the covariance matrix of the respective fit and are shown only when larger than the symbol size.}
\label{fig:aging_pNIPAM}
\end{figure}

We show in Fig.~\ref{fig:aging_pNIPAM} various dynamical quantities measured using the PCI90 COLIS camera ($q=22.2 \,\mu\si{\meter}^{-1}$), for a same suspension of pNIPAM microgels, quenched from a fluid state at $T=30$°C to either a supercooled, equilibrium state (Fig.~\ref{fig:aging_pNIPAM}a-c, $T=28$°C, cooling rate $\dot T = 0.5~\mathrm{Kmin}^{-1}$) or to a glassy state (Fig.~\ref{fig:aging_pNIPAM}d-f, $T=25$°C, $\dot T = 0.0238~\mathrm{Kmin}^{-1}$). %
For the supercooled sample, the intensity correlation functions exhibit a distinctive two-step decay (Fig.~\ref{fig:aging_pNIPAM}a), with a faster mode, related to the particles' motion in the cage formed by their neighbors~\cite{Hunterphysicscolloidalglass2012}, followed by a slower structural relaxation. Correlation functions measured at various ages, i.e. times $t_w$ since reaching the target temperature, overlap almost perfectly, suggesting that the dynamics are nearly independent of $t_w$ and the sample is almost equilibrated. To better quantify the evolution of the dynamics, we fit the final decay of $g_2-1$ with a stretched exponential, or Kohlrausch-Williams-Watts, function (lines in Fig.\ref{fig:aging_pNIPAM}a), a form widely used in the glass literature~\cite{williams1970non}:
\begin{equation}
    g_2(q,\tau)-1 = A_q\exp \left[ - 2\left( \frac{\tau}{\tau_r}\right)^{\beta} \right] +B \,.
\label{Eq:KWW}
\end{equation}
Here, $\tau_r$ is the relaxation time, $\beta$ the stretching exponent and $B\approx 0$ a baseline. $A_q$ accounts for both the speckle-to-pixel size ratio, as in Eq.~\ref{eq:g2Brownian}, and the relative weight of the structural relaxation mode.
The relaxation time issued from the fits is shown in Fig.~\ref{fig:aging_pNIPAM}b: it is longer than 100 s, more than $10^5$ times slower than in the $\varphi \rightarrow 0$ limit, and exhibits only a mild increase with $t_w$, consistently with equilibration. The stretching exponent is almost constant ($\beta = 0.358 \pm 0.005$ averaged over the full duration of the experiment), with no systematic evolution. Similarly low values of $\beta$ have been reported by Philippe \textit{et al.}~\cite{philippe2018glass} for the supercooled regime of pNIPAM microgels obtained with a different synthesis protocol. Both the nearly constant $\tau_r$ and the low $\beta$ value strongly suggest that the sample is in a nearly equilibrated supercooled state. This is further confirmed by inspecting temporal fluctuations of the dynamics, a hallmark of the dynamics of glassy systems~\cite{berthier_dynamical_2011}, which we find to be small for this sample. Dynamical heterogeneity is quantified by the dynamical susceptibility $\chi_4(\tau) \equiv <c_I(t,\tau)^2>_t - <c_I(t,\tau)>^2_t$, which we correct for the contribution due to the finite number of camera pixels as detailed in Ref.~\cite{duri_timeresolvedcorrelation_2005}, see also SI. The dynamic susceptibility is shown in Fig.~\ref{fig:aging_pNIPAM}c: it exhibits a broad peak around $100\,$s, reflecting the mild slowing down of the dynamics. The order of magnitude of $\chi_4$ is $10^{-5}$, compatible with that reported by Philippe \textit{et al.} for similar microgels; crucially, it is very small compared to $10^{-3}$, the typical value reported for strongly heterogeneous dynamics~\cite{duri_timeresolvedcorrelation_2005,dallari2020microscopic}, thus confirming that this sample is nearly equilibrated.

The scenario is very different for the sample cooled at lower $T=25\,^\circ$C, corresponding to a larger $\varphi$. Figure~\ref{fig:aging_pNIPAM}d) shows representative autocorrelation functions for various $t_w$. The separation between the fast decay and the slow structural relaxation is more marked than for the sample at $T=28.00\,^\circ$C, and aging is more pronounced, as highlighted in Fig.~\ref{fig:aging_pNIPAM}e. Two aging regimes may be distinguished. Initially ($t_w \le 3\times 10^4\,$s), $\tau_r$ increases by more than a factor of two. After this strong aging, the system enters a dynamical state characterized by an overall mild aging, but with large fluctuations of the relaxation time (gray shaded region in Fig.~\ref{fig:aging_pNIPAM}e). These differences are accompanied by an evolution of the stretching exponent, which increases from $\beta = 0.541 \pm 0.003$ for $t_w \le 3600$~s to $\beta = 0.63 \pm 0.01$ during the last hour of the experiment. %
Dynamic heterogeneity in the second regime is well captured by $\chi_4$, Fig.~\ref{fig:aging_pNIPAM}f. The dynamic susceptibility shows a very pronounced increase for long delays, hinting at a peak for $\tau \approx  2\times  10^4~\mathrm{s}$. For glassy systems, $\chi_4$ typically exhibits a peak on a timescale roughly corresponding to the system average relaxation time~\cite{berthier_dynamical_2011}. Here, unfortunately, we do not have access to these long timescales, since the measurement lasted less than a full relaxation of $g_2-1$ (see Fig.~\ref{fig:aging_pNIPAM}d). However, the strong increase of $\chi_4$ up to values about one decade larger than for the supercooled sample at $T=28\,^\circ$C, the aging behavior, and the marked separation of the cage dynamics and structural relaxation time all confirm that the sample at $T=25\,^\circ$C is in a glassy, out-of-equilibrium state.

\subsubsection{Dynamics driven by NIR laser heating  }
\label{sec:glass_driven}

\begin{figure}
\centering
\includegraphics[width=0.95\textwidth]{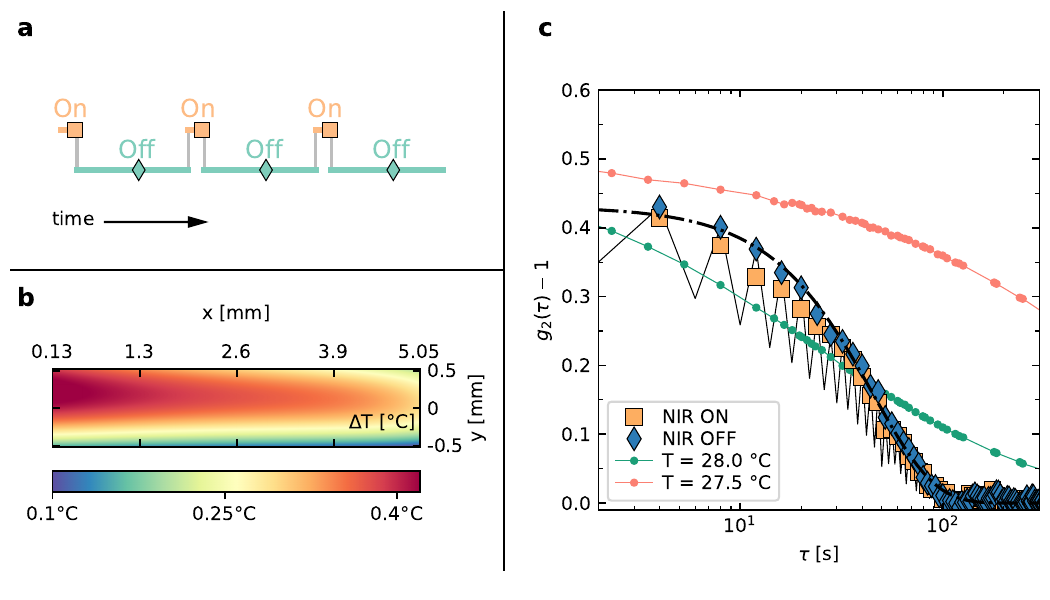}

\caption{Dynamics induced in a thermosensitive colloidal glass by periodic heating. a): temporal scheme of the NIR laser ON-OFF cycles. The duration of the ON and OFF states is 0.5 s and 3.5 s, respectively. The symbols indicate the time at which PCI90 images are acquired. b): Local temperature increase during the ON status. 
c): Intensity correlation function for a pNIPAM glass prepared at $T= 27~^{\circ}$C, measured at q=$22.2 \,\mu\si{\meter}^{-1}$). Symbols: $g_2-1$ obtained by correlating images taken while the NIR has the same status (ON or OFF, as shown by the legend). The two correlation functions are very similar, only the result of a fit with Eq.~\ref{Eq:KWW} of the OFF status is shown for clarity (dot-dashed line). Black solid line: total correlation function calculated starting from an OFF configuration. Green and orange line and small symbols: correlation functions measured with no NIR laser perturbation, for $T= 28~^{\circ}$C and $T= 27.5~^{\circ}$C, respectively. 
}
\label{fig:NIR_g2}
\end{figure}

We leverage the thermo-sensitivity of pNIPAM microgels coupled to the NIR laser of COLIS to demonstrate thermally-driven dynamics in a colloidal glass. The same pNIPAM suspension as in Sec.~\ref{sec:glass_spontaneous} is brought to a glassy state by cooling the sample from  $T=30.00\,^\circ$C to $T=27.00\,^\circ$C at a cooling rate of $1~\mathrm{Kmin}^{-1}$. In the glassy state, the NIR laser is used to impose a cyclic perturbation, as shown in Fig.~\ref{fig:NIR_g2}a. In the ON state, the NIR laser power is %
$300$~mW; the duty cycles is 12.5\%, to reduce the overall heat injected in the sample while allowing for a significant $T$ increase during the ON state. %
We probe the effect of the local heating due to the NIR laser on the glassy dynamics by acquiring speckle images with the PCI90 camera ($q=22.2~\mu\mathrm{m}^{-1}$), using a stroboscopic scheme, in the same spirit as in the so-called echo protocol for cyclically sheared samples~\cite{hebraud_yielding_1997,hohler_periodic_1997}: two images are collected for each cycle, respectively at the end of the ON status and in the middle of the OFF one, see Fig.~\ref{fig:NIR_g2}a.

Figure~\ref{fig:NIR_g2}b shows $\Delta T(\mathbf{r})$, the local temperature increase during the ON status with respect to the OFF status. The map has been calculated by averaging 10 speckle images for each status, following the procedure introduced in Sec.~\ref{sec:NIR_tests}. The average $\Delta T$ is $\sim0.4~^{\circ}$C, starting from $T= 27.4~^{\circ}$C in the OFF state. %
Interestingly, the temperature map is qualitatively different from that for continuous NIR heating, shown in Fig.~\ref{fig:Tmap}b. This difference stems from the heat diffusion timescale, whose order of magnitude is $d^2/\alpha \sim 10 \,\si{\s}$, with $d\sim 1\, \si{\mm}$ the typical length-scale and $\alpha \simeq 0.15\, \si{\mm^2\s}^{-1}$ the thermal diffusivity of water. Since the duration of the ON status is much shorter than the time needed for heat to be dissipated, we observe here an almost `instantaneous' temperature profile which is marginally influenced by the boundary conditions imposed by the thermalized cell walls. In Fig.~\ref{fig:NIR_g2}b, the NIR laser enters the sample from the left and is attenuated during propagation, due to absorption. This is clearly reflected in the temperature map, with a hotter region where the laser enters and a cooler one on the exit side.

The stroboscopic acquisition scheme adopted here reveals the effects of the NIR cycles on the sample dynamics. In Fig.~\ref{fig:NIR_g2}c, we show $g_2-1$ functions computed correlating images taken during only the ON or OFF status (yellow squares and blue diamonds, respectively). The two correlation functions are very similar; a fit with Eq.~(\ref{Eq:KWW}) yields $\tau_r = 76\pm 1\,$s and $\beta = 1.33 \pm 0.04$ for the ON status, and $\tau_r = 73\pm 1\,$s and $\beta= 1.44 \pm 0.04$ for the OFF status (dot-dashed line in Fig.~\ref{fig:NIR_g2}c). By contrast, if the correlation function is calculated for all the acquired images (both ON and OFF, black solid line), the curve shows an oscillating pattern, with lower values when correlating configurations taken at different $T$. This behavior is reminiscent of that of soft solids under oscillating shear~\cite{aime2023unified,edera_deformation_2021}. For a small number of imposed cycles ($t<10\,$s), the cycles of swelling and de-swelling of the pNIPAM microgels play a role similar to that of small-amplitude periodic shear, with the sample almost returning to its previous microscopic state after each cycle. For larger number of applied cycles, microscopic rearrangements add up, eventually leading to the full decay of $g_2-1$.
The analogy with shear-induced plasticity is also reinforced by the shape of the stroboscopic correlation functions. Here, we find $\beta > 1$ for a modest perturbation (about 20\% decorrelation between successive ON-OFF statuses), similarly to the low-shear regime of ~\cite{aime2023unified}. 

There are however obvious differences in the kind of perturbation imposed to the sample: oscillations of the local volume fraction in COLIS, \textit{vs} a global shear in mechanical tests. In particular, one may wonder if here the loss of correlation is due to spontaneous relaxations occurring in the sample when its effective volume fraction decreases, due to the NIR-induced heating. To address this question, we compare the dynamics during the NIR cycles to the spontaneous dynamics measured with no NIR beam at two relevant temperatures: $T= 27.5~^{\circ}$C, close to the average sample $T$ in the OFF status, and, as a limiting case, $T=28.0~^{\circ}$C, the maximum local $T$ observed in the ON status. Figure~\ref{fig:NIR_g2}c) shows that the $g_2-1$ functions at rest have both a longer decay time and a more stretched shape as compared to those measured while cycling the NIR laser, thus confirming that the latter induces additional plastic rearrangements, distinct from the spontaneous dynamics.

\section{Discussion and conclusions} \label{sec:conclusions}

The ground tests performed on COLIS have validated the numerous functionalities of the setup. Dynamic light scattering measurements, both polarized and depolarized, yielded translational and rotational diffusion coefficients of Brownian particles fully consistent with expectations from the particle specifications and from independent measurements on conventional laboratory setups. Furthermore, PCI and DLS data have been shown to be fully consistent with each other. A peculiar feature observed for PCI170 and DLS170 is the low-angle scattering contribution due to particles illuminated by a small fraction of the incident beam that is back-reflected at the SALS exit windows. The magnitude of this  contribution depends on the refractive index of the solvent, which determines the reflection coefficient at the sample-window interface, and the relative magnitude of the forward scattered intensity compared to back-scattering. This effect can be mitigated in laboratory setups, e.g. by immersing the sample cell in a larger  index-matching vat. Here, it cannot be avoided, due to the design constraints inherent to a flight apparatus, as well as the need to include an optical window for collecting SALS data.

We demonstrated both static and dynamic SALS on diluted colloidal particles. SALS data are notoriously affected by stray light: when subtracting an optical background measured with the cell filled by the solvent, the static $I(q)$ was successfully measured down to scattering vectors as small as $0.4~\si{\micro m}^{-1}$, while $I(q)$ was overestimated by at most a factor of 2 at smaller $q$, due to imperfect background correction. Dynamic SALS was successfully demonstrated down to $\sim 0.5~\si{\micro m}^{-1}$, data at lower $q$ being affected by stray light and residual sedimentation and convection. This test was particularly stringent, since no optical background correction was implemented, and the sample was particularly sensitive to stray light, since the particles used in the test gave a SALS signal smaller than that expected for typical flight samples, such as colloidal gels.

The experiments on protein crystallization demonstrated that cell D of COLIS allows for very fast and reproducible $T$ quenches with minimal $T$ undershoot. DLS performed in short runs, coupled to absorbance measurements, was successfully used to follow in detail the formation of nucleation precursors in lysozyme solutions and to measure crystallization induction times, with results fully consistent with previous works. In particular, DLS was able to resolve both larger clusters and individual small protein molecules, with $R_h = 2.4~\si{nm}$.

The use of an auxiliary NIR laser beam to locally heat the scattering volume, for water-based samples, and space-resolved light scattering (PCI) are among the most innovative features of COLIS. We leveraged the space resolution afforded by PCI and the change of scattering intensity with temperature of a concentrated suspension of pNIPAM to establish maps of the local heating induced by the NIR beam. This approach allowed us to measure the transient response to a time-varying NIR illumination, as well the $T$ map in the steady state. Note that the latter would not have been accessible using the approach of previous works, based on the $T$ dependent emission of fluorophores dissolved in water~\cite{ross2001temperature,ruzzi2023optothermal}, because of convection motion that would have continuously stirred the solvent. The dense pNIPAM suspensions used here, by contrast, should be regarded as an effective porous medium with nanometer-sized pores, which strongly hamper water advection. The NIR experiments have furthermore showed that the dynamics of glassy samples may be significantly sped up with swelling-deswelling cyclces, paving the way for investigating the yielding transition in an original driving mode, complementary to the extensively studied mechanical shear~\cite{divoux_ductile--brittle_2024}. In the absence of NIR driving, we were able to measure glassy relaxations and dynamic heterogeneity  on time scales up to several $10^5~\si{s}$, demonstrating the excellent thermal and mechanical stability of COLIS. 

The numerous measurement lines, the use of detectors with complementary features, e.g. the relatively slow, multi-element CMOS cameras and the fast, single-element APDs, the blending of scattering and imaging capabilities, and the advanced thermal control of the sample environment, including with the NIR laser, make COLIS a unique apparatus for investigating the structure and dynamics of soft matter in microgravity. Although COLIS has been primarily designed for measuring weakly scattering systems in Fourier space, its flexible design allows for extensions to other samples and experimental configurations. Examples that may be of interest to the community include using the SALS camera and the PCI170 and DLS170 lines for Diffusing Wave Spectroscopy~\cite{weitz_diffusing-wave_1993} measurements on strongly multiple scattering samples, in the transmission and backscattering geometry, respectively, or using COLIS as a low-magnification multi-view imaging system, through the PCI lines.

\section{Methods}
\subsection{Materials} \label{sec:materials}

Brownian particles for the DLS measurements of Sec.~\ref{sec:Brownian_DLS} were nearly monodisperse polystyrene (PS) particles of diameter 277 nm (Invitrogen) suspended in a 80/20 w/w mixture of glycerol and water at temperature $T = 25$°C, with refractive index $n = 1.444$ and nominal viscosity of 0.045%
~Pa s, yielding a nominal diffusion coefficient of 0.042~$\si{\micro m~s}^{-1}$. The particle volume fraction $\varphi$ was lower than $10^{-4}$.\\

For DDLS, Sec.~\ref{sec:Brownian_depolarized}), we used a $\varphi = 10^{-2}$ aqueous suspension of Hyflon MFA, a copolymer of tetrafluoroethylene (TFE) and perfluoromethylvinylether (PF-MVE), produced by Solvay-Solexis S.p.A., Bollate, Italy. The colloids are spherical particles with an average radius of $R = 90\pm 2$ nm  and a polydispersity of 4\%,  as determined by DLS on a laboratory-based conventional setup. They have a low refractive index ($n_p=1.352$), allowing for index matching for single scattering investigations at arbitrary $\varphi$. Furthermore, MFA particles can be roughly pictured as a collection
of polymer crystallites, embedded into an amorphous matrix. As
a consequence, the particles are optically anisotropic, thus allowing for DDLS measurements.
The same particles, but at higher $\varphi = 0.2$, where used to form the colloidal gels studied in Sec.~\ref{sec:gel}. To screen electrostatic repulsions, approximately 100 mM of NaCl were added to the suspension.
Triton X100, a non-ionic surfactant,  was used at a volume fraction $\phi_T=0.5$, both as a steric stabilizer and as a depletant agent~\cite{Buzzaccaro2007}, to induce attractive interactions~\cite{AsakuraInteractionBodiesImmersed1954}. Finally, about 8 \% ww of urea was added to the suspension to fully match the particle refractive index at room temperature. Note that in this system the strength of the depletion interactions may be changed by varying temperature, thus providing a convenient way to induce gelation or redisperse the suspension~\cite{Degiorgio1996, Buzzaccaro2007, Piazza2011}. 

For testing SALS, Sec.~\ref{sec:SALS_results}, we used a suspension of silica (SiO$_2$) particles for static light scattering and a suspension of PS particles for dynamic light scattering. The SiO$_2$ particles (microParticles Gmbh) 
had a diameter of 2.01 $\mu\mathrm{m}$ and refractive index $n_p = 1.42$; they were suspended in MilliQ water, at $\varphi = 2 \times 10^{-5}$. The PS particles (Invitrogen CML Latex) had a radius of 230 nm and where suspended at a volume fraction of $8.4 \times 10^{-5}$ in a 78.2/21.8 w/w mixture of water and glycerol at a temperature $T = 25$°C, with refractive index $n \approx 1.36$ and nominal viscosity of 1.6~mPa s, yielding a nominal diffusion coefficient of 0.59~$\mu\mathrm{m^2s}^{-1}$. The particles and solvent density were nearly matched, $\rho_{PS} \approx \rho_{solvent} \approx 1.05~\mathrm{g cm}^{-3}$, to minimize sedimentation effects.

The colloidal glasses used in Secs.~\ref{sec:NIR_tests}
and~\ref{sec:glass} are the same as those to be used for the experiments onboard the ISS. They are water-based, concentrated suspensions of poly-N-isopropylacrylamide (pNIPAM)
microgel particles with ultralow crosslinking density~\cite{brugnoni_synthesis_2019, scotti_phase_2020}, which reduces their scattering cross section thereby
making them suitable for single scattering experiments. In the range 18°C
to 30°C, the particle radius $R$ decreases with temperature, due to the
change of affinity of the polymer chains with the solvent~\cite{pelton_preparation_1986}. For our microgels, $R = 130\,$nm at $T=25\,^\circ$C, as measured by DLS. The mass concentration was 34 g/L, corresponding to an effective volume fraction of 1.30 at $T=25$°C [See e.g.~\cite{truzzolillo_bulk_2015} for how the effective volume fraction is calculated from the known mass fraction; note that effective volume fractions larger than one are possible for deformable, interpenetrable particles such as our microgels.]

For the protein experiments of Sec.~\ref{sec:protein}, lysozyme stock solutions were prepared by dissolving 7 mg/mL lyophilized lysozyme (PanReac Applichem 232-620-4) in a sodium acetate buffer (50 mM, pH 4.5). The pH was adjusted with HCl to give the desired pH 4.50. The solutions were filtered two times (0.2 µm filters) to remove all undissolved protein particles  and dialyzed two times against the buffer to remove the excess salt of the lyophilized lysozyme. For each dialysis step, 100 ml of solution was dialyzed in 2 l of buffer for 24 h. The membranes used for the dialysis were Spectra/Por 3 with a 3.5kD cut off. After dialysis, the concentration of lysozyme was determined by spectrophotometry. For a concentration measurement, the stock solution is dissolved (dilution factor 200 to 500) and absorbance was measured at 280 nm using a CARY 100 Bio spectrophotometer and an absorption coefficient of e=2.64 ml/mg was used to determine the concentration. The precipitant stock solutions of NaCl solutions were prepared separately by dissolving the appropriate amount of NaCl in the same buffer and Sodium Azide was added in order to neutralize any biological contamination. After complete dissolution of NaCl and NaN3, all solutions are filtered to 0.2 µm to remove any undissolved material. The lysozyme and NaCl stock solution are finally mixed to obtain the final concentrations. Typical concentrations of lysozyme, NaCl and NaN$_3$ in the mixed samples are 7 mg/ml, 65-70mg/mL and 0.5 mg/ml, respectively. 

\section{Data availability statement}
All data shown in the figures are available on Zenodo [link to be provided]

%

\input{biblio.bbl}%

\section*{Acknowledgments} \label{sec:acknowledgments}
We thank A. Scotti, A. Petrunin, and W. Richtering for providing us with the pNIPAM microgels, R. Elancheliyan and D. Truzzolillo for their characterization, and E. Lichaire for help in some measurements. We are grateful to ESA for  continuously supporting the Colloidal Solids project. We thank D. Tomuta for critically reading the manuscript.

\section*{Author Contribution} 

Overall design of COLIS: RP, LC.
Software for operating COLIS and onboard data analysis: EL, LC.
Conducting experiments and data analysis: AM, SB, QG, JB, NS, EL, YS, DM, JL, LC.
Writing the initial version of the manuscript: AM, SB, QG, JL, LC.
Critical reading of the manuscript: all authors.
Final edit of the manuscript: LC.

\section*{Funding} 
The work of QG, DM and JL was partially supported by the European Space Agency (ESA) and the Belgian Federal Science Policy Office (BELSPO) in the framework of the PRODEX Programme,
Contract No. ESA AO-2004-070. AM is a postdoctoral researcher supported by the French CNES (Centre National d’Etudes Spatiales). LC gratefully acknowledges support from CNES and the Institut Universitaire de France. This research was partly funded by The Italian Ministry of University and Research (PRIN Project 20222MCY75) and by the Italian Space Agency (ASI, Project ``NESTEX'').

\section*{Competing interests statement} 
The authors declare no conflict of interests.

\end{document}


\title{\centering{Supplementary Information to:\\}COLIS: an advanced light scattering apparatus for investigating the structure and dynamics of soft matter onboard the International Space Station}

\maketitle
\section*{Table of contents:}
\begin{itemize}
    \item{Measuring temperature maps}
    \begin{itemize}
        \item Figure S1: Scattered intensity map from PCI images.
        \item Figure S2: Calculation of the pixel-dependent temperature map using  calibration curves.
    \end{itemize}

    \item{Correction of the four-point susceptibility $\chi_4$ for the contribution of finite ROI size}
    \begin{itemize}
        \item Figure S3: Extrapolation of $\chi_4$ to the infinite ROI size limit.
    \end{itemize}
    
    \item{Additional Dynamic Light Scattering data for the Brownian particles of Sec. 3.1.1 of the main text}
    \begin{itemize}
        \item Figure S4: PCI45, PCI90, and DLS90 intensity correlation functions.
    \end{itemize}
    
\end{itemize}

\newpage
\section{Measuring temperature maps}
In order to monitor locally the sample temperature while operating the NIR laser with a pNIPAM microgel sample, a spatially- and temporally-resolved approach is needed. We adopt a strategy involving measuring the scattered intensity in the field of view of the PCI90 camera. Since the scattered intensity of a dense suspension of pPNIPAM microgels strongly depends on temperature, changes in the sample temperature are quantified by measuring intensity changes in the PCI images.\\
\begin{figure}[ht]
    \centering
    \includegraphics[width=1\linewidth]{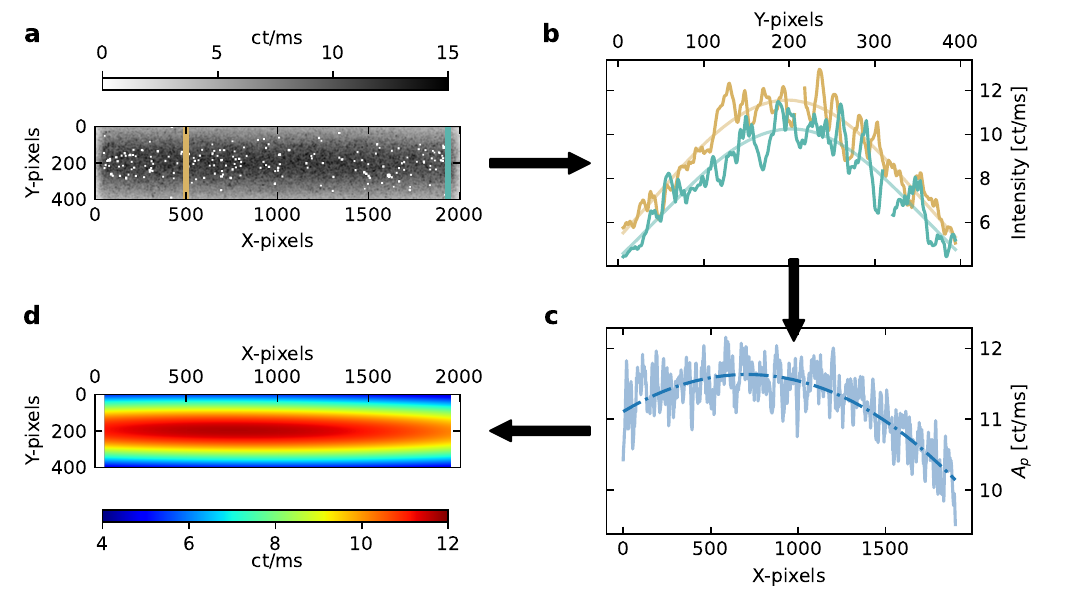}
    \caption{Procedure to establish a spatially-resolved calibration curve relating the scattered intensity to the local temperature. The arrows indicate the sequence of steps for processing the data. a) PCI90 image of the average scattered intensity for a pNIPAM sample at $T=27~^\circ$C. The image is the average of $\sim 2000$ frames acquired over $\sim$5.5 hours. Note that speckles are still partially visible, due to the slow dynamics of pNIPAM. b) Intensity level along two representative column of pixels of the image in a), indicated using the same color code in both panels. The curves in b) are fitted with a Gaussian function to obtain a smooth intensity profile, disregarding short-distance fluctuations due to the residual speckles. c) Amplitude $A_p$ of the Gaussian fit as a function of the $X$ coordinate, see  text for details. The blue dashed curve is a 2$^{\text{nd}}$-order polynomial fit to the data. d) Reconstructed spatial distribution of the scattered intensity, where speckle fluctuations have been fully smoothed.}
    \label{S1}
\end{figure}

As a first step, we establish a spatial map of the average scattered intensity at a fixed, uniform temperature $T$ (no NIR beam), where spatial variations of $I$ are due to the non-uniformity of the illuminating Gaussian beam. For several temperatures in the range $T_{min}=18\,^\circ$C to $T_{max}=30\,^\circ$C we acquire sequences of the scattered intensity. Each sequence contains up to a few thousands of frames, which are averaged to reduce pixel-to-pixel intensity fluctuations due to speckles. `Hot' spots, i.e. regions with anomalously high scattering intensity due to spurious contributions such as dust or aggregates in the sample are masked and the average image is convoluted with a square kernel box of side 10 pixels, to further smooth the intensity. An example of the result is shown in Fig.(\ref{S1}a), where the laser beam entering from the left side of the image is clearly visible. The intensity profile along each column of the image is then fitted with a Gaussian function:
\begin{equation}
    I_X(Y) = A_p(X) \exp[-(Y-Y_0(X))^2/4\sigma^2(X)]
\end{equation}
with $A_p$ the amplitude, $Y_0$ the center and $\sigma^2$ the variance of the Gaussian, respectively. $X$ indicates the position of the column of pixels. The Gaussian fit is an effective and robust way to model the local intensity, removing the residual effect of speckles and scattering impurities. Figure (\ref{S1}b) shows an example of the Gaussian fit, while Fig. (\ref{S1}c) shows the $X$ dependence of $A_p$. Overall, the beam is slightly attenuated while passing across the sample, since scattering removes power from the propagating beam. We attribute the slight decrease of $A_p$, and thus $I$, near $X=0$ (entrance wall) to limitations of the collection optics. Finally, in order to further smooth the intensity profile, the fit parameters $A_p$, $Y_0$, and $\sigma^2$ are themselves fitted to polynomials in $X$ of degree 2, 1, and 0, respectively. As an example, Fig.(\ref{S1}c) shows the parabolic fit of $A_p(X)$.
The polynomials are used to reconstruct a smooth map of the scattered intensity, as shown in Fig.(\ref{S1}d). We emphasize that the smoothing and fitting procedure described above is essential in order to suppress intensity fluctuations due to speckles and to obtain reliable results in the subsequent calibration step described below.

\begin{figure}[ht]
    \centering
    \includegraphics[width=0.8\linewidth]{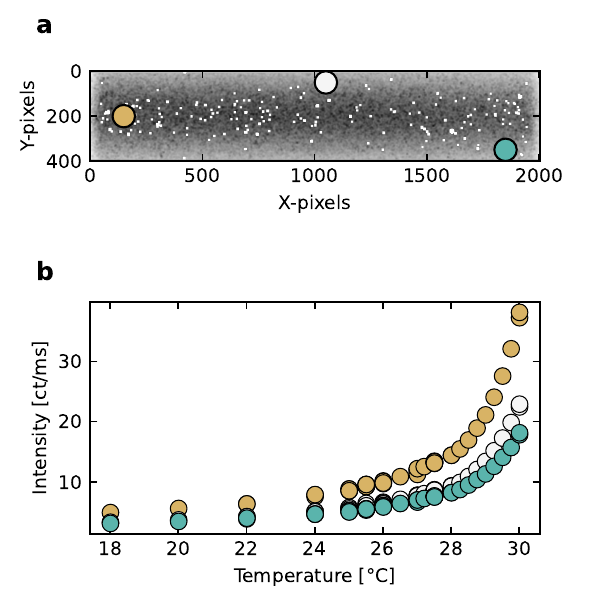}
    \caption{a) Same average image as in Fig.(\ref{S1}), with highlighted the 3 points where we extracted the calibration curves shown in the bottom panel. b) Calibration curves (scattered intensity as a function of temperature) for the points shows in the upper panel.}
    \label{S2}
\end{figure}

The mapping procedure has been repeated for more than 50 times in the temperature range of interest, yielding, for each image point with coordinates $(X,Y)$, a calibration curve $I$ \textit{vs} $T$, as shown in Fig.(\ref{S2}). These spatially-resolved calibration curves are used for inferring the local temperature while operating the NIR laser, by reading, for each image pixel, the temperature corresponding to a given measured intensity level. Note that the calibration curves are built using intensity levels normalized with respect to the exposure time, to allow for optimizing the exposure time in any given experiment.

\newpage
\section{Correction of the four-point susceptibility $\chi_4$ for the contribution of finite ROI size}

\begin{figure}[ht]
    \centering
    \includegraphics[width=0.8\linewidth]{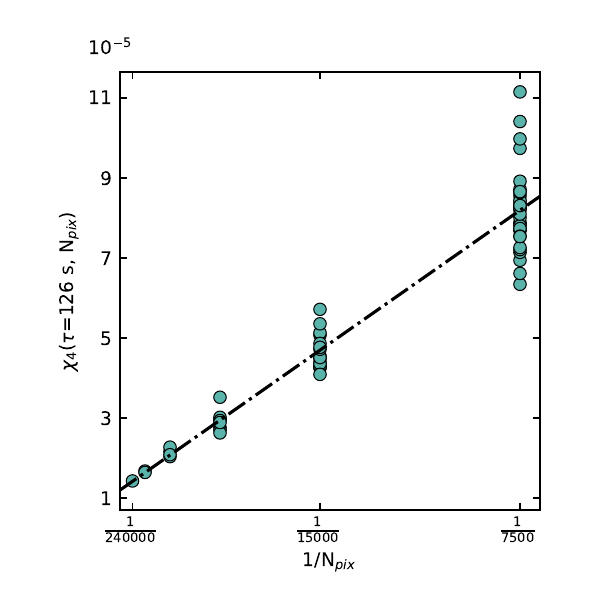}
    \caption{Four-point susceptibility at $\Delta t=126\,$s calculated for different number of pixels for the glassy sample reported in the main text (Fig.9). Black-dashed line is the best linear fit to the data.}
    \label{S3}
\end{figure}

The four-point susceptibility $\chi_4$ \cite{Berthier2011,Berthier2011_2} is calculated as the variance of the degree of correlation, see Sec. 3.6.1 of the main text:
\begin{equation}
\chi_4(\tau) = \langle c_I(t,\tau)^2 \rangle_t -  \langle c_I(t,\tau)  \rangle_t^2.
\end{equation}
As discussed in~\cite{Duri2005}, the finite size of the ROI over which
$c_I$ is calculated introduces a statistical noise that adds up to the real value of $\chi_4$. This additional noise scales as $1/N_{pix}$, with $N_{pix}$ the number of pixels belonging to the ROI. It is thus clear that in the limit of $N_{pix}\rightarrow\infty$, one recovers the desired, noise-free $\chi_4$ value. To estimate the correct value of $\chi_4$, we follow an extrapolation procedure similar to that of \cite{Duri2005}: we calculate $\chi_4$ for ROIs of different size, plot the result vs $1/N_{pix}$ and linearly extrapolate to $1/N_{pix} \rightarrow 0$.

In PCI, however, different ROIs correspond to different scattering volumes and in principle the dynamics could be heterogeneous across the scattering volume. Thus, the extrapolation procedure needs to be somehow modified. We start by choosing an initial region of interest, ROI$_N$, containing $N$ pixels. We then extract $N/N_{pix}$ sub-ROIs, randomly choosing at each time $N_{pix}$ pixels from the initial ROI$_N$. This ensures that, even with a reduced number of pixels, we still sample the entire volume corresponding to the initial. Figure (\ref{S3}) shows the result of such procedure for the glassy sample discussed in Sec. 3.6.1 of the main text. An initial ROI$_N$ with $N= 240000$ pixels is selected and is then divided into $2^m$ sub-ROIs, with $m=[0,1,2,3,4,5]$. The dynamic susceptibility is calculated for each sub-ROI, and the $\chi_4$ \textit{vs} $1/N_{pix}$ data are fitted with a straight line (dashed line). The intercept of the fitted line corresponds to the desired $\chi_4(\tau)$ value in the $N_{pix}\rightarrow\infty$ limit. This procedure is repeated for each delay $\tau$, leading to the four-point susceptibilities shown in Figs. 9c,f of the main text.

\section{Additional Dynamic Light Scattering data for the Brownian particles of Sec. 3.1.1 of the main text}
\begin{figure}[ht]
    \centering
    \includegraphics[width=1\linewidth]{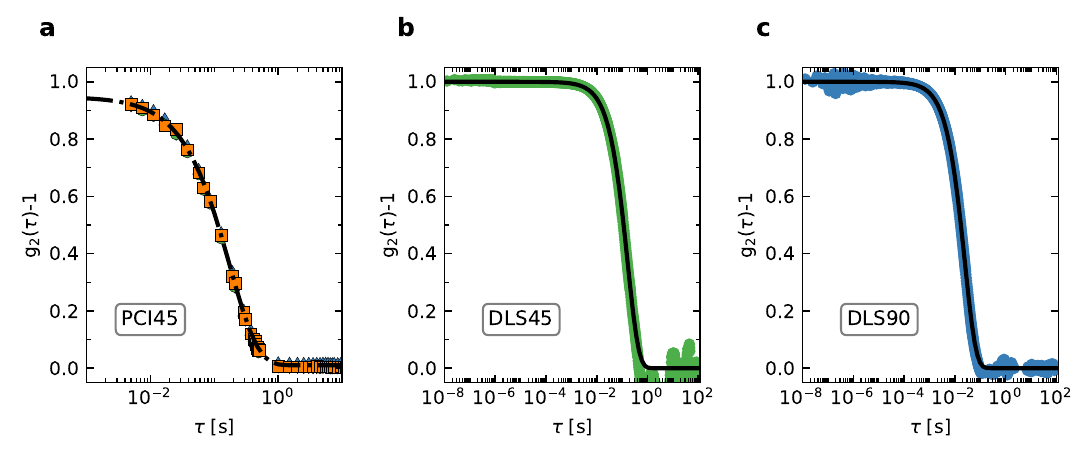}
    \caption{Dynamics of Brownian particles measured by polarized DLS. a) intensity correlation function measured by the PCI45 camera. b) Intensity correlation function measured by the DLS45 correlator. For a) and b), the scattering angle is $\theta = 29.32^{\circ}$. c) Intensity correlation function measured by the DLS90 correlator.
    }
    \label{S4}
\end{figure}
\newpage
\bibliographystyle{Science}